\documentclass{aastex631}

\usepackage{color}

\shorttitle{Energetic Particles from QSLs and CSs at the Sun}
\shortauthors{Schwadron et al.}
\graphicspath{{./}{figures/}}

\begin{document}

\title{Energetic Particles from Quasi-Separatrix Layers and Current Sheets at the Sun}

\correspondingauthor{Nathan Schwadron}
\email{Nathan.Schwadron@unh.edu}

\author[0000-0002-3737-9283]{Nathan A. Schwadron}
\affiliation{University of New Hampshire, Morse Hall, 8 College Road, Durham, NH, 03824, USA}

\author[0000-0002-2633-4290]{Ronald M. Caplan}
\affiliation{Predictive Science Inc., 9990 Mesa Rim Road, Suite 170, San Diego, CA, 92121, USA}

\author[0000-0003-1662-3328]{Jon A. Linker}
\affiliation{Predictive Science Inc., 9990 Mesa Rim Road, Suite 170, San Diego, CA, 92121, USA}

\author[0000-0001-6590-3479]{Erika Palmerio}
\affiliation{Predictive Science Inc., 9990 Mesa Rim Road, Suite 170, San Diego, CA, 92121, USA}

\author[0000-0003-2124-7814]{Matthew A. Young}
\affiliation{University of New Hampshire, Morse Hall, 8 College Road, Durham, NH, 03824, USA}

\begin{abstract}
Quasi-separatrix layers (QSLs) at the Sun are created in regions where channels of open magnetic flux have footpoints near regions of large-scale closed magnetic flux. These regions show rapid changes in curvature and field-strength. Numerical simulations of a relaxed coronal magnetic field and solar wind  using the Magnetohydrodynamic Algorithm outside a Sphere (MAS) model coupled to the Energetic Particle Radiation Environment Module (EPREM) model indicate common sources of energetic particles over broad longitudinal distributions in the background solar wind. These regions accelerate energetic particles from  QSLs and current sheets. Here, we develop an analytical framework to describe the acceleration of energetic particles due to the magnetic field changes within and near separatrix layers. The reduced field strength near the separatrix layer drives magnetic field magnitude changes  that  accelerate energetic particles in the presence of plasma flow along the structure. Separatrix layers are prone to magnetic reconnection, creating fluctuations in the field that propagate out from the Sun, and release material previously contained within closed magnetic field structures, which are often rich in heavy ions and $^3$He ions. Thus, we present a model of energetic particles accelerated from separatrix layers in the corona. Our results provide a plausible source for seed populations near the Sun. 
\end{abstract}

\keywords{}


\section{Introduction} \label{sec:intro}

Solar energetic particles (SEPs) are high-energy, charged particles that are accelerated and transported in the solar corona and solar wind.  Large fluxes of SEPs are present in solar particle events (SPEs), which represent a significant hazard for humans and technological infrastructure. SPEs can harm aircraft avionics, communications, and navigation systems. They also represent a possible risk to the health of airline crews and passengers on polar flights. In space, SPEs can be hazardous for crews of Low Earth Orbit spacecraft and the International Space Station, especially when engaged in extravehicular activity.   They
may also imperil crews of future manned lunar or interplanetary missions. Understanding the origin of SPEs and predicting the resulting fluxes at different locations in the heliosphere is therefore not only significant scientifically, but is also necessary from a space weather perspective.

SEPs are associated with solar eruptions; specifically solar flares and coronal mass ejections  \citep[CMEs, e.g.,][]{reames2013}.  Since the 1990s, SEPs have been roughly divided into impulsive events, associated with magnetic reconnection in short-duration solar flares, and gradual events, believed to be accelerated by shock waves driven by CMEs \citep{Reames:1999}.   An important question for shock-accelerated SEPs in large gradual events is the source of the suprathermal seed population, which may be remnants of earlier flares/CMEs \citep{mewaldtetal2012}.  A more complex picture has emerged from recent observations \citep[][and references therein]{anastasiadisetal2019}, with reconnection possibly playing a role in gradual events.  In large solar eruptions, CMEs and flares are typically closely associated and are believed to be the result of the same underlying process that  disrupts and reconfigures the coronal magnetic field \citep[e.g.,][]{forbes2000}.  While the initiation mechanism(s) are still under debate, magnetic reconnection is seen to be an important part of the energy release process in CMEs \citep[e.g.,][and references therein]{greenetal2018}.  

Solar flares are thought to be fundamentally related to magnetic reconnection. In the standard model of flare reconnection, open magnetic field lines are swept through a current sheet, reconnecting with each other to form cusp-shaped loops in the corona. These cusp-shaped
loops then relax into a more potential, rounded state \cite[]{Priest:2002}. This is sometimes referred to as field line shrinkage \cite[]{Svestka:1987}. 
One of the difficulties in this model for
understanding SEPs is that the contraction of the
field lines that leads to particle acceleration \cite[]{Somov:1997, Tsuneta:1998} occurs on closed magnetic field
lines. Therefore, the release of SEPs on open magnetic field lines
requires either interchange magnetic reconnection between open
magnetic field lines and the previously closed magnetic loops or cross
field diffusion from the closed magnetic field lines onto open
magnetic field lines.
The outflow exhausts from any form of magnetic reconnection can be
very fast, provided that the Alfv\'en speed in the outflow region is
large. In the corona, Alfv\'en speeds are typically 500~km/s in weak
field regions, but can become extremely large (10,000--20,000~km/s)
where fields are strong. Shock waves will
likely exist at the termination of the exhaust. Shock waves produced
by impulsively-driven reconnection may be important during flares or
during the emergence of magnetic flux from the photosphere into the
corona. \cite{Forbes:1986}, \cite{Blackman:1994}, and \cite{Workman:2011} have investigated such shock waves by studying numerical
experiments using 2D magnetohydrodynamics (MHD). 



For particles accelerated by reconnection processes to be measured in interplanetary space, they must escape the the closed magnetic field region.  Interchange reconnection between open and closed field lines is considered the most likely mechanism.  This process 
was considered originally as a
concept for the origin of slow solar wind.  \cite{Axford:1977b} and 
\cite{Fisk:2001} argued that the solar wind
from the polar coronal holes should be substantially different from the
solar wind from the low-latitude return region where open magnetic
flux encounters and undergoes reconnection with closed coronal
loops. The resulting solar wind should therefore be highly variable, as is
observed, and presumably slow. This concept also offers
an explanation for compositional differences between the fast and slow
solar wind \cite[]{Schwadron:1999}. 
The material stored in loops may, as a result of wave heating, contain enhancements in elements with low first ionization potential (FIP).
Further, the loops
may act as conduits for the solar wind by temporarily storing and
heating plasma to higher coronal temperatures.

The reconnection process between open and closed magnetic field causes
intermittent changes in the footpoints of open magnetic flux tubes. After  open and closed magnetic flux tubes reconnect, the open and closed field footpoints exchange their locations.  

\cite{Crooker:2002} demonstrated that CMEs should also cause
reconnection between closed and open magnetic flux.  Until 1995,
disconnection at the Sun was thought to be the only solution to the
problem of balancing the magnetic flux of CMEs
added to the heliosphere, in spite of the fact that the expected solar
wind signature of disconnection was rare. Disconnection was pictured
as merging between open field lines to create completely disconnected
U-shaped structures or merging between closed field lines to create,
in two dimensions, completely disconnected plasmoids. Since 1995, both
theoretical and observational studies have made important
contributions toward a revision of this solution. On the basis of a
synthesis of these studies, \cite{Crooker:2002} suggested that the
primary flux balancing mechanism is not complete disconnection but
rather merging between closed and open fields, i.e, 
interchange reconnection.  

\begin{figure}[th!]
\centering
\includegraphics[width = 0.85\textwidth]{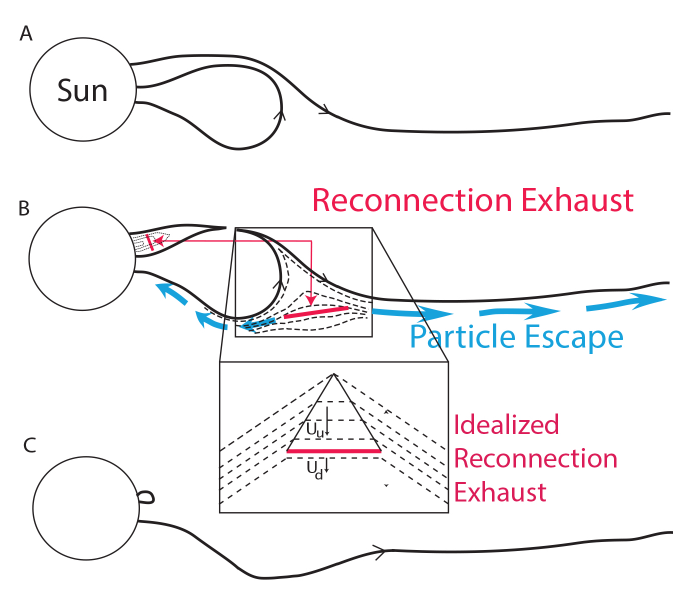}
\caption{Schematic diagram of interchange reconnection and acceleration of
energetic particles  near the reconnection exhaust termination. An open magnetic field and a closed magnetic field line approach one another (Panel A). The
simple closed topology shown here may actually be more complex, as discussed by \cite{Gosling:1995}. If the closed and open magnetic flux reconnect (Panel B), there are two outflow
exhausts from the reconnection site. One exhaust is directed into the closed field region and
the other exhaust is directed into the open field. The energetic particles accelerated within the exhaust that forms on the open magnetic field line (the lower side of panel B) are
free to move throughout the inner heliosphere, and may provide seed populations for particles
accelerated at the bow shocks of coronal mass ejections. Panel C shows the reconfiguration of the open and
closed magnetic flux away from the reconnection exhaust. The inset in Panel B shows an idealized reconnection exhaust created by Alfv\'en wings as fast-mode  waves propagate out and away from the reconnection site. The reconnection exhaust forms where the cusp-shaped
structures  relax into a more potential state. }
\label{fig:recon}
\end{figure}

The closed CME magnetic flux undergoing interchange reconnection will
move the reconnected open flux by at least the CME footpoint
separation distance (see Figure \ref{fig:recon}). Since the polarity of CME footpoints tends to
follow a pattern similar to the Hale cycle of sunspot polarity,
repeated CME eruption and subsequent reconnection will naturally
result in latitudinal transport of open solar flux.

The topology in 3D magnetic configurations \cite[]{Baum:1980}  is known to play a significant role in eruptive processes leading to solar flares and CMEs. Current sheets form naturally near thin separatrices where the magnetic field vanishes or becomes very weak \cite[e.g.,][]{Aly:1990, Low:1988, Lau:1993}. Magnetic  reconnection along a separator changes the current system, often leading to the large release of energy stored within them   \cite[]{Henoux:1987}. Observed solar flares show that the field lines closest to the separator connect to bright ribbons in the chromosphere \cite[]{Somov:1998},  indicating the relationship between large current sheets and energy release from the magnetic field and plasma. 

Quasi-separatrix layers \cite[QSLs; ][]{priest:1995, Demoulin:1996one, Demoulin:1996three}  are defined as regions of high squashing factor $Q$ \cite[]{titov:2007}, where $Q$ measures geometric distortion and connectivity gradients. Large magnetic field gradients are favorable regions of electric current density accumulation (resulting from shear between flux systems) and are therefore locations susceptible to magnetic reconnection \cite[]{syrovatskii:1981, priest:1995, masson:2009}.  Observed flares have been studied based on the QSL field connectivity down to the chromosphere \cite[]{Demoulin:1997, Mandrini:1997, Bagala:2000}. There have also been applications that have associated QSL reconnection at open–closed flux system boundaries (interchange reconnection) to the origin and evolution of the slow solar wind \cite[e.g.,][]{antiochosetal2011, linker_etal-2011-evolutionopenmagnetic, crooker:2012, lynch:2023}. 

MHD models of the solar corona \citep{antiochosetal2011, linker_etal-2011-evolutionopenmagnetic, Titov:2011} revealed that a web of separators and QSLs, primarily associated with the helmet streamer belt, should surround the Sun in the middle corona.  This ``S-web'' was proposed as a natural location for interchange reconnection and a source for the slow solar wind.  Interchange reconnection dynamics were subsequently demonstrated in MHD simulations \citep{higginsonetal2017,higgensonlynch2018}, and observational evidence for the S-web as a dynamical slow solar wind source has recently been presented  \citep{bakeretal2023,chittaetal2023}.\textcolor{black}{The S-web appears in plots of coronal magnetic-field polarity, typically displayed in \textit{signed log Q} format, defined as S-log Q $\equiv \mathrm{sign}(B_{r})\,\mathrm{log}[Q/2 + (Q^{2}/4 - 1)^{1/2} ]$ \citep[e.g.,][]{titov11}.}

In this paper we develop an analytic framework to describe energetic particle acceleration above large-scale current sheets and QSLs, associated with the S-web, where energetic particle populations are accelerated.  The results of  analytical treatments developed here suggest that QSL-associated particle acceleration naturally provides the seed population for energetic particles. 

The paper is organized as follows:   We first consider the effects of separatrix layers on energetic particles in \S 2, and develop an analytical differential equation  to describe the acceleration  process based on the Focused Transport Equation.  In \S 3 we describe modeling of separatrix layers using the SPE Threat Assessment Tool (STAT). In \S 4 we develop an analytic solution to explain particle acceleration from separatrix layers, and we connect the analytic solution to the numerical simulations discussed in \S 3.  In \S \ref{sec:random}, we connect the cascade of magnetic energy down to small-scales from separatrix layers with particle acceleration as a superposition of random stochastic processes.  Subsequently, we apply the results of these models to describe key characteristics of seed populations from separatrix layers (\S\ref{sec:qsFluxes}), and the composition of these seed populations (\S\ref{sec:composition}). In \S\ref{sec:conclusion}, we conclude the paper, listing the key aspects of seed populations that follow from particle acceleration from QSLs.  The paper includes an appendix with a full derivation of the analytic solution for acceleration at separatrix layers.  

\section{Particle Acceleration at Separatrix Layers}
\label{sec:qslEnergeticParticlesEffects} 
\begin{figure}[th!]
\centering
\includegraphics[width = 0.85\textwidth]{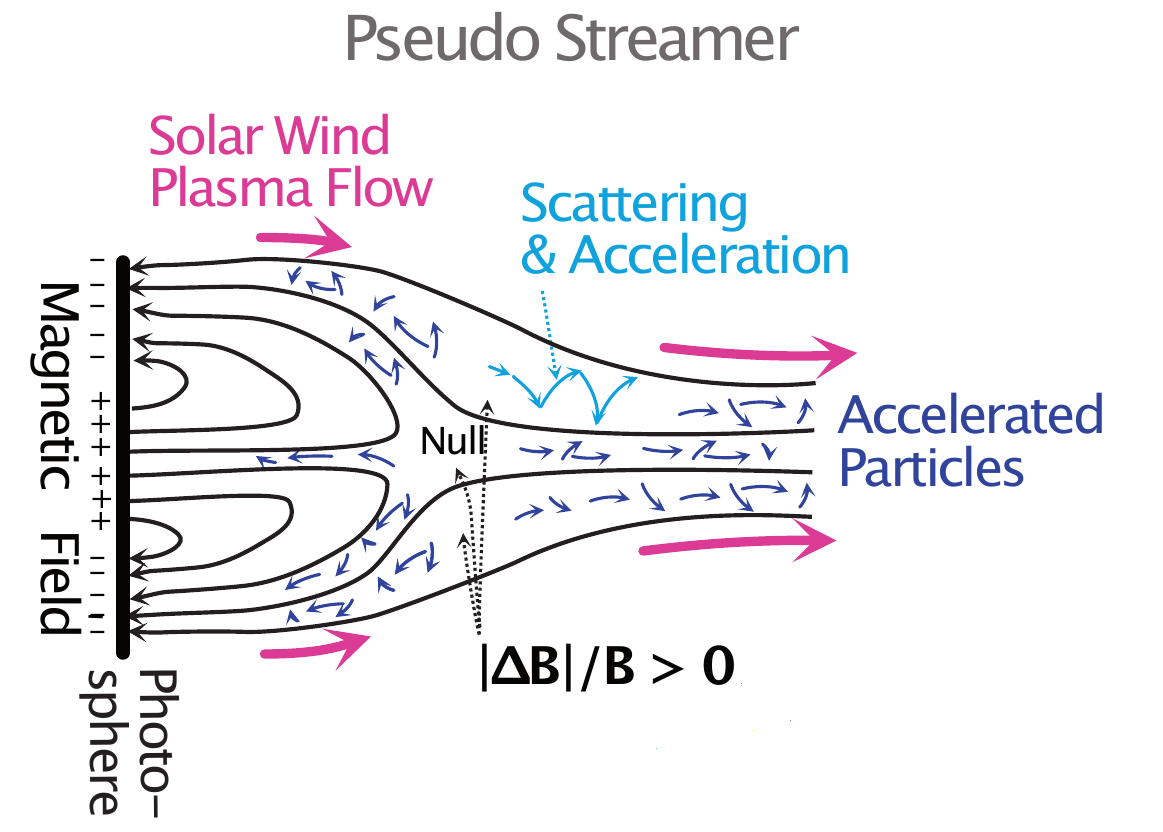}
\caption{Schematic diagram of particle acceleration near an idealized pseudo-streamer, a form of a QSL. The energetic particles accelerated near the QSL experience strong gradients in the magnetic field strength on the open magnetic field lines. Particles moving through this structure  with plasma flow undergo particle acceleration as they are scattered through the magnitude changes in field strength. Particles both gain and lose energy through the interaction, and the presence of scattering introduces a randomization of the events causing diffusion in momentum space. 
 }
\label{fig:illustration}
\end{figure}

STAT simulations (see Section \ref{sec:model}) have revealed that outflows from QSLs accelerate particles, predominantly where strong gradients in the magnetic field strength are accompanied by a strong plasma flow along the magnetic field lines. The QSL associated with a pseudo-streamer is illustrated in Figure~\ref{fig:illustration}.

The particle acceleration process in the separatrix layer involves three significant components. First, the separatrix layer introduces strong gradients in the field strength along open magnetic structures. The reduction in field strength occurs near nulls in the field where there are rapid changes in field structure, often resulting from the transition between predominantly closed field structures and open field lines. QSLs typically do not result from polarity changes on the open field lines. Instead these result from changes in field topology and complexity, often in regions where sinews of open field impinge on a hierarchy of closed field structures. The illustrated structure in Figure~\ref{fig:illustration} represents an idealized QSL where the transition from closed to open field structures is apparent in conjunction with the field null immediately above the closed field loops. 

The second component is the existence of plasma flow on the open field structures. This  solar wind plasma accelerates along field lines in the low corona. While the particle motion through a static field structure can do no work on the particle, the plasma itself is in motion. As a result, particles experience a change in the magnetic field strength with time.    A good example of this effect is for a particle with 90$^\circ$ pitch-angle, which  convects with the plasma flow through the field structure on the open field line. The particle first experiences  a depression in the magnetic field strength, causing the particle energy to drop in direct proportion to the reduction in field strength. As the particle convects further through the structure, it experiences an increase in field strength, causing the particle energy to increase. Without scattering, there is no net  change to the particle energy  after it has convected fully through the separatrix layer. However, as we discuss next, the presence of scattering enables diffusion in energy. 

The third factor is scattering, which introduces random statistical behavior. As an extreme case, consider a particle with speed $v_0$, energy $E_0 = m v_0^2/2$, and 0$^\circ$ pitch-angle. We take the field line to have field strength $B = B_0$ outside the separatrix layer, and a field strength reduction of $\Delta B$ within the structure. The particle travels into the separatrix layer with $0^\circ$ pitch-angle, and therefore experiences no energy change. Within the separatrix layer, where the field strength is reduced $B = B_0 -\Delta B$,  the particle is scattered to 90$^\circ$ pitch-angle and then convects with the plasma flow through the structure and back to the magnetic field with strength $B = B_0$ outside the separatrix layer. As the particle propagates with the plasma, it experiences a changing field strength, and the first adiabatic invariant is conserved. When the particle emerges from the separatrix layer, it has a higher energy, $ E_1  = E_0 B_0 /(B_0 - \Delta B)$. The work in changing the particle energy is done not by the magnetic field, but by the plasma in moving the particle through the field structure. 

The opposite interaction is also possible. For example, a  particle convected into the separatrix layer at 90$^\circ$ pitch-angle, experiences loss of energy. Within the separatrix layer, the particle can be scattered to 0$^\circ$ pitch-angle and thereby propagate out of the separatrix layer at lower energy, $E_1 = E_0 (B_0 - \Delta B)/ B_0$. 

While these examples are extreme cases, they demonstrate that scattering fundamentally changes the interaction with the separatrix layer. The scattering randomly causes either increases or decreases in particle energy. The result of many random scattering interactions in the separatrix layer is the diffusion of particles in momentum space. Since more particles enter with lower momentum (and therefore, lower energy), this diffusion in momentum-space yields a net increase in energy throughout the particle distribution. This process is essentially a magnetic pump, encountered typically in energetic particle physics when magnitude variations, or magnetosonic waves, in the magnetic field are present \cite[]{Schwadron:1996}. The  process is described as transit-time damping since the waves or turbulence associated with the field variations are strongly damped in the process of particle acceleration.  

One primary difference between the separatrix layer acceleration and transit-time damping is that the changes in magnetic field strength are associated with the structure of the magnetic field, not waves or turbulence. The existence of plasma flow through the  structure is also critical since a static field fundamentally cannot do work on the particles.  Lastly, since waves or turbulence in the field introduce time variations, it is the wave or turbulence field that is eroded through transit time damping. In contrast, the separatrix layer is a feature of the  magnetic field structure, and it is the plasma flow that does work on the particles as the distribution is convected through the structured field variations.   

A QSL is  a special case of the introduction of separatrix nulls associated with topology, complexity, and structural changes in the magnetic field. Current sheets in the open field typically accompany the same field reductions above the nulls at the separatrices between closed  and open field structures. The mechanism formalized in this paper  is associated with the open field lines at current sheets and QSLs near the Sun where particles are pumped magnetically within the accelerating solar wind.

We isolate the acceleration effect by first considering the Focused Transport Equation in its entirety:
\begin{eqnarray}
\frac{d f}{d t} & + & v \mu \hat{e}_b \cdot \nabla f + \frac{1-\mu^2}{2} \left [  - v \hat{e}_b \cdot \nabla \ln B - \frac{2}{v} \hat{e}_b \cdot \frac{d\mathbf{u}}{dt} 
+ \mu \frac{d \ln n^2/B^3 }{dt} \right] \frac{\partial f}{\partial \mu}  \nonumber \\
& + & \left[ -\frac{\mu }{v}\hat{e}_b \cdot \frac{d \mathbf{u}}{dt} + \mu^2 \frac{d \ln n/ B}{d t} + \frac{1-\mu^2}{2} \frac{d \ln B}{dt}  \right] \frac{\partial f}{\partial \ln p} \nonumber \\
& = & \frac{\partial}{\partial \mu} \left( D_{\mu\mu} \frac{\partial f}{\partial \mu} \right)\,,
\label{eq:focused_transport}
\end{eqnarray}
\noindent where $f$ is the particle distribution function, $v$ is the particle speed, $\mu$ is the cosine of the particle's magnetic pitch angle, $\hat{e}_{b}$ is a unit vector parallel to the magnetic field, $B$ is the magnitude of the magnetic field, $\mathbf{u}$ is the solar-wind flow velocity, $n$ is the solar-wind density, $p$ is the particle momentum, and $D_{\mu\mu}$ is the diffusive scattering coefficient.

Separatrix layers are consistent with a slow down in the flow along the flux tube. Evaluation of acceleration terms will be presented in \S \ref{sec:analytic}, and indicates the dominance of terms involving $d \ln B/dt $, with some comparatively modest changes in $d \ln n/dt$. Based on this analysis, we ignore terms proportional to changes in the flow speed, $d\mathbf{u}/dt$, and algebraically arrange the remaining terms to arrive at the following approximation to the Focused Transport Equation: 
\begin{eqnarray}
\frac{d f}{d t} & + & v \mu \hat{e}_b \cdot \nabla f - \frac{1-\mu^2}{2} \left [  v \hat{e}_b \cdot \nabla \ln B  + 3 \mu \frac{d \ln (B/n^{2/3}) }{dt} \right ] \frac{\partial f}{\partial \mu}  \nonumber \\
& + & \left[  \frac{1}{3}\frac{d \ln n}{d t}  - \frac{3 \mu^2 - 1}{2} \frac{d \ln (B/n^{2/3})}{d t}  \right] \frac{\partial f}{\partial \ln p} \nonumber \\
& = & \frac{\partial}{\partial \mu} \left( D_{\mu\mu} \frac{\partial f}{\partial \mu} \right) \,.
\label{eq:focusedTransport}
\end{eqnarray}
For the scattering coefficient, we use the following \cite[]{Schwadron:1994}: 
\begin{eqnarray}
D_{\mu \mu} = \frac{1-\mu^2}{2 \tau_s}\,,
\end{eqnarray}
\noindent where $\tau_s^{-1}$ is the scattering rate, and $\tau_s$ is related to the parallel scattering mean free path, $\lambda_{\parallel} = \tau_s v$. 

The approximation made is that pitch-angle diffusion drives the distribution toward isotropy. 
We now take moments of the distribution function assuming a truncated Legendre polynomial expansion about the unperturbed distribution function, $f_{0}$, and as a function of $\mu$:
\begin{eqnarray}
f \approx f_0 P_0 + f_1 P_1(\mu) + f_2 P_2(\mu)\,,
\end{eqnarray}
\noindent where $P_i$ are Legendre Polynomials
\begin{eqnarray}
P_0 & = & 1 \nonumber \,, \\
P_1 & = & \mu \nonumber \,,\\
P_2 & = &  (3 \mu^2 -1)/2 \nonumber  \,.
\end{eqnarray}
Retaining the first three moments of the distribution function is equivalent to a third-order expansion. This approach has become standard for deriving the Parker equation and higher-order expansions that capture shear-related quantities in addition to compression \cite[]{Williams:1993}.  
We take the $0^\mathrm{th}$-moment of equation~(\ref{eq:focusedTransport}) by integrating over the equation by $1/2 \int_{-1}^1 d\mu $, resulting in the following:
\begin{eqnarray}
\frac{df_0}{dt} + \frac{v}{3} \hat{e}_b \cdot \nabla f_1 - \frac{v}{3}  \left( \hat{e}_b \cdot \nabla \ln B \right) f_1 - \frac{3}{5} \frac{d\ln (B/n^{2/3})}{dt} f_2 &  - &  \frac{1}{5} \frac{d \ln (B/n^{2/3})}{dt} \frac{\partial f_2}{\partial \ln p} \nonumber \\  & + &  \frac{1}{3} \frac{d \ln n}{dt} \frac{\partial f_0}{\partial \ln p} = 0 \,. 
\label{eq:zerothMoment}
\end{eqnarray}
Similarly, we take the $1^\mathrm{st}$-moment of equation~(\ref{eq:focusedTransport}) by integrating over the equation by $3/2 \int_{-1}^1 d\mu \mu$, resulting in the following:
\begin{eqnarray}
\frac{df_1}{dt} + v \hat{e}_b \cdot \nabla \left( f_0 + \frac{2}{5}f_2 \right)  -  \frac{3v}{5}  \left( \hat{e}_b \cdot \nabla \ln B \right) f_2 & -  & \frac{3}{5} \frac{d\ln (B/n^{2/3})}{dt} f_1  -  \frac{2}{5} \frac{d \ln (B/n^{2/3})}{dt} \frac{\partial f_1}{\partial \ln p} \nonumber \\ &  + &  \frac{1}{3}\frac{d\ln n}{dt}\frac{\partial f_1}{\partial\ln p} = -\frac{ f_1}{\tau_s} . 
\end{eqnarray}
We take the $2^\mathrm{nd}$-moment of equation~(\ref{eq:focusedTransport}) by integrating over the equation by $5/2 \int_{-1}^1 d\mu P_2(\mu)$, resulting in the following:
\begin{eqnarray}
\frac{df_2}{dt} +   \frac{2v}{3} \hat{e}_b \cdot \nabla f_1 + \frac{v}{3}  \hat{e}_b \cdot \nabla \ln B f_1 & - &  \frac{3}{7} \frac{d\ln (B/n^{2/3})}{dt} f_2 -  \frac{d \ln (B/n^{2/3})}{dt} \frac{\partial f_0}{\partial \ln p}  \nonumber \\
&  - &  \frac{2}{7} \frac{d \ln (B/n^{2/3})}{dt} \frac{\partial f_2}{\partial \ln p}   +   \frac{1}{3}\frac{d\ln n}{dt}\frac{\partial f_2}{\partial\ln p} = -\frac{3 f_2}{\tau_s} . 
\end{eqnarray}
Following conventional perturbation theory, we expand this equation in terms of a smallness parameter $\epsilon \sim \tau_s f_0^\prime$ where $f_0^\prime$ denotes the time differential $\partial f_0/\partial t$, spatial differential $v \nabla f_0$, or other rates multiplied by terms involving $f_0$. In other words, the expansion is about the scattering time, which defines the smallest timescale of the quantities considered. This leads to following approximations for $f_1$ and $f_2$:
\begin{eqnarray}
f_1  & \approx &  - \lambda_\parallel \hat{e}_b \cdot \nabla f_0 \\
f_2 & \approx & \frac{\tau_s }{3} \frac{d\ln (B/n^{2/3})}{dt } \frac{\partial f_0}{\partial \ln p}
\end{eqnarray} 
where $\lambda_\parallel = \tau_s v$ is the parallel scattering mean free path. 
Substituting these terms into the zeroth-moment equation~(\ref{eq:zerothMoment}) yields:
\begin{eqnarray}
\frac{df_0}{dt} - \nabla\cdot \left( \bar{\bar{\kappa}} \cdot \nabla f_0 \right) - \frac{\nabla \cdot \mathbf{u}}{3}  \frac{\partial f_0}{\partial \ln p} -  \frac{1}{p^2} \frac{\partial }{\partial p} \left( p^2 D_{pp} \frac{\partial f_0}{\partial p} \right) = 0 . 
\label{eq:transport}
\end{eqnarray}
where 
\begin{eqnarray}
\bar{\bar{\kappa}} =  \frac{\lambda_\parallel v}{3} \hat{e}_b \hat{e}_b\,, \qquad 
\frac{D_{pp}}{p^2} =  \frac{\tau_s}{15} \left[\frac{d \ln (B/n^{2/3})}{dt} \right]^2  \end{eqnarray}
and we have substituted $-\nabla\cdot\mathbf{u}$ for $d\ln n/dt$ according to the continuity equation.
Therefore, the effect of the separatrix layer is to create localized particle acceleration due to diffusion in momentum space. The diffusion coefficient resembles that derived from transit-time damping magnetic field fluctuations in field magnitude \cite[]{Schwadron:1996}. 

The form of acceleration derived here implies that particles experience energy changes through multiple interactions with the separatrix region. Particles passing directly through the separatrix region without scattering experience little or no change in energy after exiting the region, and without scattering, these particles cannot return to the  separatrix region for further interactions. Scattering, therefore, is a critical part of the acceleration process, both for interrupting the adiabatic change process within the reconnection region, and for enabling particles to interact with the reconnection region over multiple encounters. 

There are two significant particle acceleration terms in the Focused Transport Equation:
\begin{eqnarray}
\left[ \frac{1}{3} \frac{d\ln n}{dt}  \right] \frac{\partial f}{\partial \ln p}\,,  \qquad \mbox{and} \qquad  \left[ \frac{(1 - 3\mu^2 )}{2} \frac{d\ln (B/n^{2/3})}{dt}  \right] \frac{\partial f}{\partial \ln p}\,.
\end{eqnarray}
The first term is associated with compression at shocks or compression regions, and the second term is associated with a combination of changes in magnetic flux density and density ($d\ln (B/n^{2/3})/dt$). 

Unlike diffusive shock acceleration, this second-order acceleration term occurs wherever there are strong gradients in the quantity $B/n^{2/3}$. Although the momentum diffusion term increases with the scattering time $\tau_s$, the diffusion rate cannot grow indefinitely since the scattering time must be sufficiently small to drive the distribution function  toward isotropy. In the limit of very large scattering times, the diffusion in momentum space is suppressed as particles propagate through the separatrix region, experiencing only local adiabatic changes in momentum and pitch-angle in response to changes in the field strength. 

In this section, we have shown that time-dependent  fluctuations in magnetic field strength and density lead to second-order momentum diffusion of suprathermal and energetic particles. We derived the effect using a Legendre polynomial expansion of the distribution function as it evolved according to the Focused Transport Equation. The derived second-order momentum diffusion is consistent with magnetic pumping in regions where there are strong gradients in the magnetic field magnitude, but the process also requires work done by the plasma  flow through the magnetic field structures. The generalization of this second-order acceleration process is required to understand the particle acceleration observed in energetic simulations at or near separatrix layers. 

\section{Modeling of Separatrix Layers: MHD Simulations Integrated with a Focused Transport Model}
\label{sec:model}

 We  show results of a numerical model of SEPs applied to the separatrix layers that form in the Sun's corona. We employ the SPE Threat Assessment Tool \citep[STAT;][]{Linker_etal-2019-CoupledMHDFocused}. STAT couples MHD simulations of CME events from Predictive Science Inc.'s Corona Heliosphere (CORHEL-CME) modeling suite \citep{Linker:2024} with focused transport simulations of solar energetic particles (SEPs) from the University of New Hampshire's Energetic Particle Radiation Environment Module (EPREM). STAT allows users to run EPREM for previously computed Magnetohydrodynamic Algorithm outside a Sphere (MAS) simulations (including those run through CORHEL-CME) of real CME events to simulate SEP events and provide diagnostics that can be compared with observations. 
Here, we use STAT with a custom run of MAS that does not include a CME event, but instead integrates the MHD equations with a fixed boundary in a time-dependent quasi-relaxation.

\subsection{MAS Simulations}
\label{sec:MAS}

The properties of compressive regions such as shocks that drive SEP acceleration depend critically on the properties of the local plasma environment. Therefore, the MHD simulation must realistically capture these properties for the time period under study. The MAS model has a long history of continued development and applications to this problem. While models with a simple energy equation can qualitatively reproduce coronal properties \citep{Mikic_Linker-1996-LargeScaleStructure,Linker_etal-1999-MagnetohydrodynamicModelingSolar,Mikic_etal-1999-MagnetohydrodynamicModelingGlobal} and are sufficient for exploring some dynamical aspects of boundary evolution \citep{linker_etal-2011-evolutionopenmagnetic}, so-called thermodynamic MHD models \citep{Lionello_etal-2009-MultispectralEmissionSun,Riley_etal-2011-GlobalMHDModeling,rileyetal2012,Downs_etal-2013-ProbingSolarMagnetic,linkeretal2017,titovetal2017,mikicetal2018} are necessary to compute the plasma density and temperature with sufficient accuracy to simulate extreme ultra-violet (EUV) and X-ray emission observed from space. In this approach, the energy equation accounts for anisotropic thermal conduction, radiative losses, and coronal heating. Inclusion of these extra physical terms is vital for obtaining realistic Alfv\'en speeds ($V_A$) and sound speeds ($C_S$). 

To model a specific time period, a full-Sun map of the photospheric magnetic field is obtained from an observatory or flux transport model and processed to create a boundary condition for the radial magnetic field  \citep[e.g.,][]{linkeretal2017}.  For this simulation, we developed a thermodynamic MHD simulation of the global corona using the procedure and equations described by \citet{Lionello_etal-2009-MultispectralEmissionSun}, but with a Wave-Turbulence-Driven (WTD) description of coronal heating \citep[e.g.,][]{mikicetal2018}.  In this thermodynamic model, the temperature at the lower boundary is set to 17,500~K, similar to the upper chromosphere, and the upper boundary is at 30~$R_\odot$, beyond the sonic and Alfv\'en critical points. 

The CORHEL-CME modeling suite \cite[]{Linker:2024} uses MAS to compute solutions in the coronal (1--30~$R_\odot$) and heliospheric (28--230~$R_\odot$) domains separately.  Coronal solutions are used to provide the inner boundary condition for the heliospheric solutions  \citep{Lionello_etal-2013-MagnetohydrodynamicSimulationsInterplanetary}.  At the present time, STAT employs only the MHD coronal domain within EPREM, and the remainder of the heliosphere is modeled with a simple spiral magnetic field created with a radially constant solar wind speed.  STAT is presently being modified to incorporate both the coronal and heliospheric solutions in the EPREM simulations; these results will be the subject of future publications.

\subsection{EPREM Focused Transport Simulations}
\label{sec:EPREM}

EPREM models energetic particle acceleration and transport using a Lagrangian system, which co-moves with the plasma.  EPREM creates a spherical shell of simulation nodes at each time step and advances each node along the MAS flow velocity, then calculates the distribution function $f_s(t, r, p, \mu)$ according to the Focused Transport Equation. EPREM uses a relaxation-time approximation for pitch-angle diffusion with the relaxation time inversely related to the pitch-angle diffusion coefficient. This treatment is applied for simplicity and significantly reduces computational cost. 

Each node advances outward with the solar wind flow and is linked to nodes on the neighboring shells. Each linked sequence of nodes defines a simulation stream representing a velocity path line---the trajectory of fluid particles.  In steady-state (i.e., in the frame rotating with the Sun) these are also streamlines;  in places where the frozen-in assumption of ideal magnetohydrodynamics (MHD) holds, these lines also represent magnetic field lines.

The advantage of solving the transport problem in the co-moving frame is that it precludes the necessity of computing spatial gradients in flow velocity, which tend to introduce numerical errors (such as extraneous cross-field diffusivity) that accumulate over many time steps. Instead, it requires the relatively simple task of computing the rates of change in plasma number density, $n$, magnetic field, $\mathbf{B}$, and flow velocity, $\mathbf{u}$, at each stream node after being moved by a timestep. This methodology is based on the approach described in \citet{Kota_etal-2005-SimulationSEPAcceleration}, which follows from the theory developed by \citet{Skilling-1971-CosmicRaysGalaxy} and \citet{Ruffolo-1995-EffectAdiabaticDeceleration}. It was used by \citet{Kozarev_etal-2013-GlobalNumericalModeling} to study time dependent effects of SEP acceleration in the low corona during CME evolution and by \citet{Schwadron_etal-2014-Synthesis3DCoronal} to model radiation doses at 1~au during a strong SPE event. In order to solve equation~(\ref{eq:focused_transport}), EPREM needs a model of $n$, $\mathbf{B}$, and $\mathbf{u}$ at each node. Simplified scenarios can use analytic forms of these plasma quantities but realistic modeling requires the use of MHD data such as that provided by CORHEL-CME.

One recent study used STAT to model the 2000 July 14 solar proton event \citep{young:2021}. This work modeled proton acceleration to GeV energies due to the passage of the CME through the low solar corona.  The simulation results compared well to GOES-8 observations,  roughly reproducing the peak event fluxes and the timing and spatial location of the energetic particle event. The model was found to accurately describe the acceleration processes in the low corona and resolved the sites of most rapid acceleration close to the Sun. Integral flux envelopes from multiple simulated observers near Earth further improve the comparison to observations and increase potential for predicting SPEs. Broken power-law fits to fluence spectra agreed with diffusive acceleration theory over the low-energy range. Over the high-energy range, they demonstrated the variability in acceleration rate during a single SPE and mirrored the inter-event variability observed in solar cycle 23 ground-level enhancements. 

\subsection{STAT Simulation of a Relaxation Event}
\label{sec:statmodel}
While STAT is designed to model CME events, we use it here to explore particle acceleration in QSL regions through a relaxation run. Modeling of the relaxation run is a complex task that involves several steps \cite[e.g.,][]{Torok:2018}. We produce a full-Sun magnetic map to create the boundary conditions for the magnetic field. The map used here is a blend of high-resolution SDO/HMI vector magnetic field data of the active region and a lower-resolution combination of synoptic maps that were used for predicting the structure of the December 14, 2020 eclipse\footnote{See \url{https://www.predsci.com/corona/dec2020eclipse}}. 

\begin{figure*}[th!]
\centering
\includegraphics[width = 0.6\textwidth]{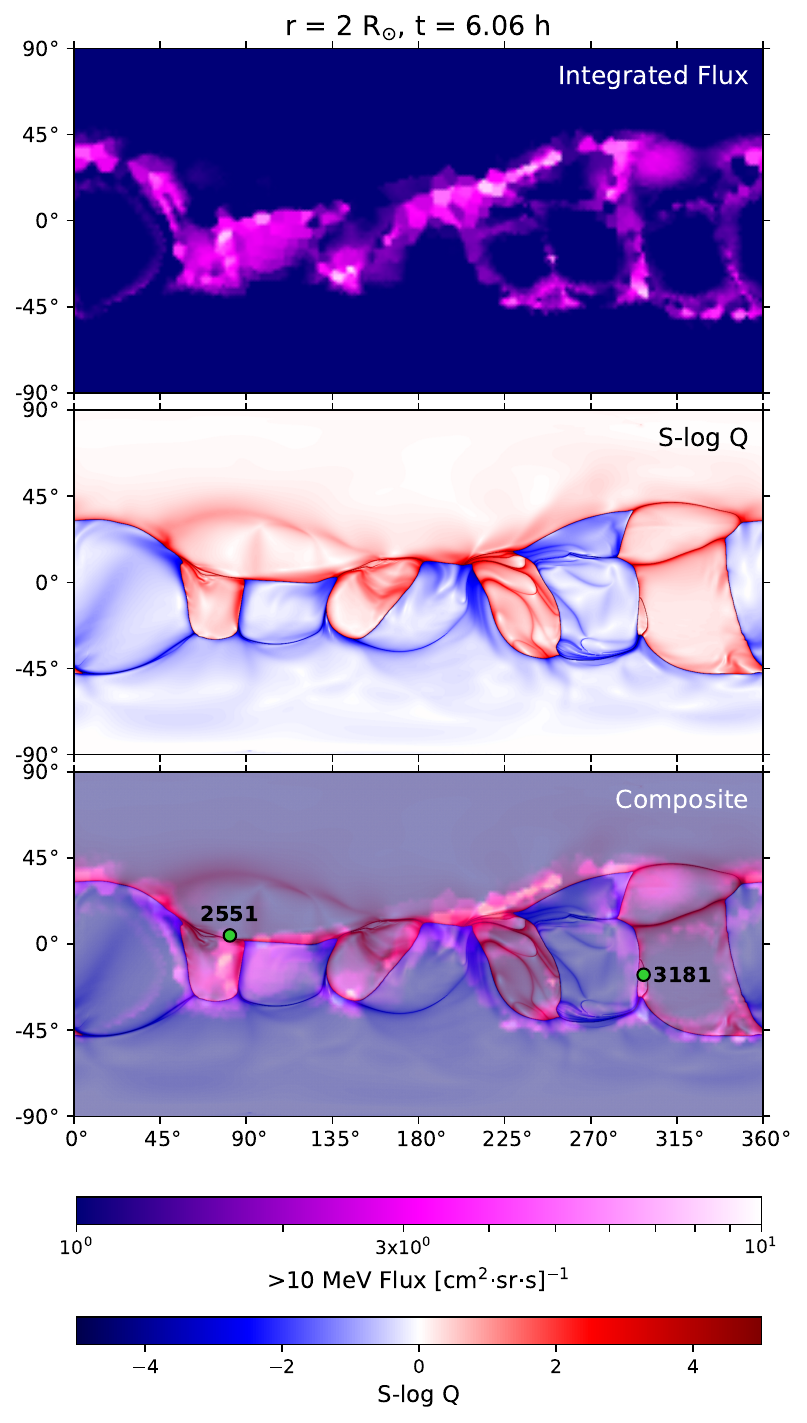}
\caption{
Distribution of energetic particle fluxes  at $2~R_{\odot}$ from the 11/29/2020 relaxation run (top panel).  The middle panel  shows the log of squashing factor $Q$ associated with QSLs (the S-web) with the red and blue coloring indicating the opposing polarity of the coronal magnetic field---data are displayed in \textit{signed log Q} format, defined as S-log Q $\equiv \mathrm{sign}(B_{r})\,\mathrm{log}[Q/2 + (Q^{2}/4 - 1)^{1/2} ]$ \citep[][]{Titov:2011}. \textcolor{black}{The heliospheric current sheet is seen as the thin boundary between the red and blue regions}. The bottom panel overlays these images to show the similarity between the spatial structures recovered. \textcolor{black}{Streams 2551, near a QSL, and 3181, near a current sheet are labeled. These streamlines are used as representative cases to describe the effects of the separatrix layer on particle acceleration.} \label{fig:relax}}
\end{figure*}

The relaxation run is created for the similar time period that of a CME that occurred on November 29, 2020 in NOAA AR 12790 as it attracted considerable attention in the solar community due to it being associated with the largest flare (M4.4 class) in three years and produced a shortwave radio blackout over the South Atlantic \citep[e.g.,][]{Kollhoff:2021, Mitchell:2021, Cohen:2021, Kouloumvakos:2022, Palmerio:2022}. The results of the relaxation run (Figure~\ref{fig:relax}) show conclusively that energetic particles are accelerated near QSLs and current boundaries over broad longitudinal regions without an emitted CME. 
 
In the simulation, the EPREM parallel diffusion coefficient depends on rigidity,  $\kappa_\parallel = \kappa_{\parallel 0} (R_g/R_{g0})^\chi$,  where $R_{g0}$ is a reference rigidity. We take $\chi = 1/3$, as was used in \cite{young:2021}, and derived from quasi-linear theory using a $5/3$ power-spectrum. The parallel diffusion coefficient is taken as $\kappa_\parallel  \propto B^{-3/4}$, where $B$ is the magnetic field strength. This dependence is slightly smaller than the $1/B$ dependence often assumed, but is roughly in-line with observations \cite[e.g.,][]{erdos:1999}



To investigate the source of the longitudinal spread in energetic particles, we traced the location of the EPREM streams producing particle fluxes at 1~au back to their origin in the low corona. We found that the source locations lie along separators, including
the heliospheric current sheet (HCS). These regions are most easily identified by plotting
the squashing factor $Q$ (the so-called S-web). The flows near separatrices locally accelerate SEPs in the simulation.

Figure~\ref{fig:relax} (top  panel)  shows the energetic particle fluxes more than 6~hr into the simulation. Significant energetic particle fluxes remain relatively close to the Sun (near 2~$R_{\odot}$) and the distribution of these fluxes strongly resembles the QSL pattern observed in the Figure's bottom panels.  
The overlays in the bottom row make the explicit association between the QSL boundaries and energetic particle fluxes. The regions of enhanced energetic particle fluxes partly follow the outline of the HCS, \textcolor{black}{which is identified as the boundary between red and blue regions in the middle panel.} There are significant areas  where the energetic particle enhancements follow the QSL boundary even outside the current sheet (e.g., peudo-streamers). Conversely, there is not a one-to-one association between regions of high $Q$ and elevated energetic particle fluxes. The particle acceleration is  most significant in regions  with strong and relatively localized reductions in the magnetic field. Surprisingly, the enhanced energetic particle fluxes do not exclusively exist in regions of strong flow compression. The vast majority of QSL and current sheet regions studied  show that the acceleration develops in response to the field reductions in the absence of strong flow compression.

\begin{figure*}[htbp!]
\centering
\includegraphics[width = 0.9\textwidth]{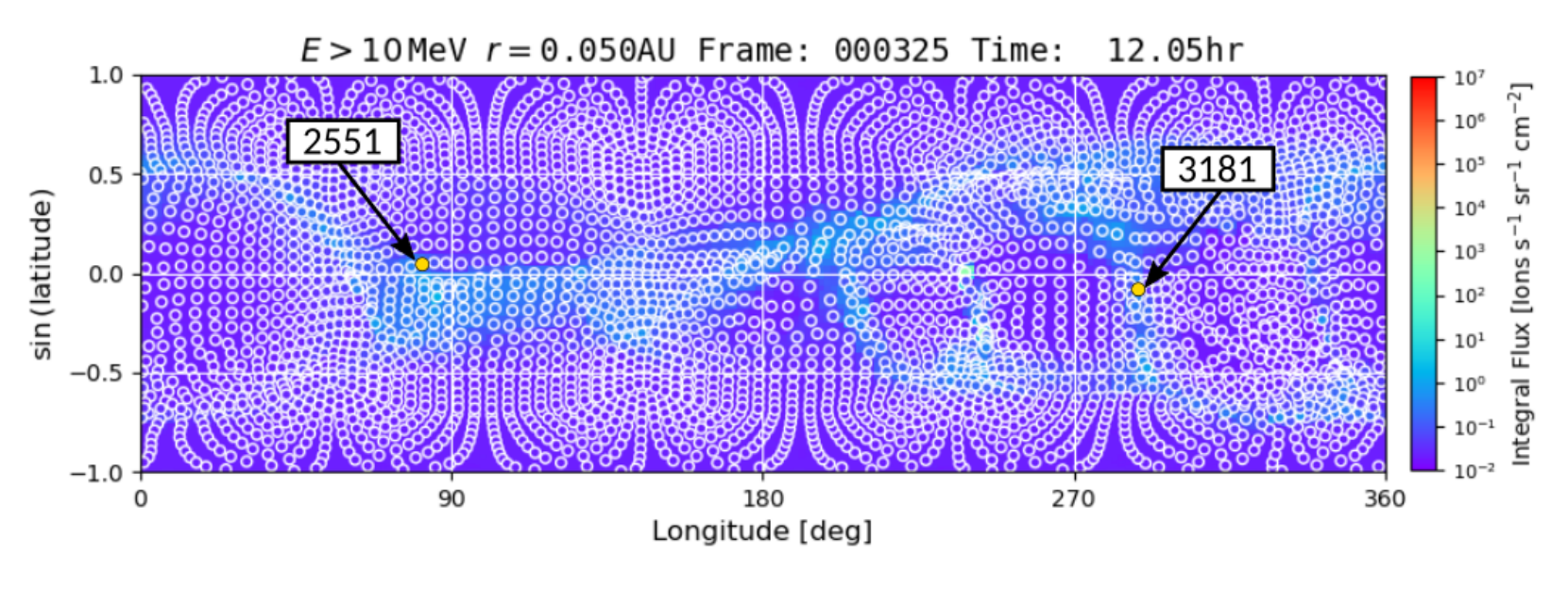}
\caption{
Integrated flux greater than 10 MeV of energetic particles at 10~$R_{\odot}$ 12~hr into the simulation period. The simulation includes the background solar wind configuration used from the November 29, 2020 event. Node locations of streamlines 2551 and 3181 are labeled. \textcolor{black}{The EPREM simulations follow nodes out through evolving plasma flow. An individual line of nodes follows a specific streamline in the simulation.  The two streamlines chosen are representative cases: streamline 2551 passes through a QSL without a current sheet, whereas streamline 3181 passes through the current sheet. There is some variation along individual streamlines, but the general behavior observed at the streamlines identified highlights the physical mechanisms responsible for particle acceleration.  } 
\label{fig:DASL}}
\end{figure*}

\textcolor{black}{The EPREM simulations follow nodes out through evolving plasma flow. An individual line of nodes follows a specific streamline in the simulation. } Figure \ref{fig:DASL} shows the locations of the EPREM streamlines at 12 hours into the simulation.  Streamlines 2551 and 3181 are highlighted, where streamline 2551 occurs at a QSL that lies away from a current sheet, whereas streamline 3181 lies along a current sheet. 
\textcolor{black}{ The two streamlines chosen are representative cases. There is some variation along individual streamlines, but the general behavior observed at the streamlines identified highlights the physical mechanisms responsible for particle acceleration.  } 

\begin{figure*}[htbp!]
\centering
\includegraphics[width = 0.7\textwidth]{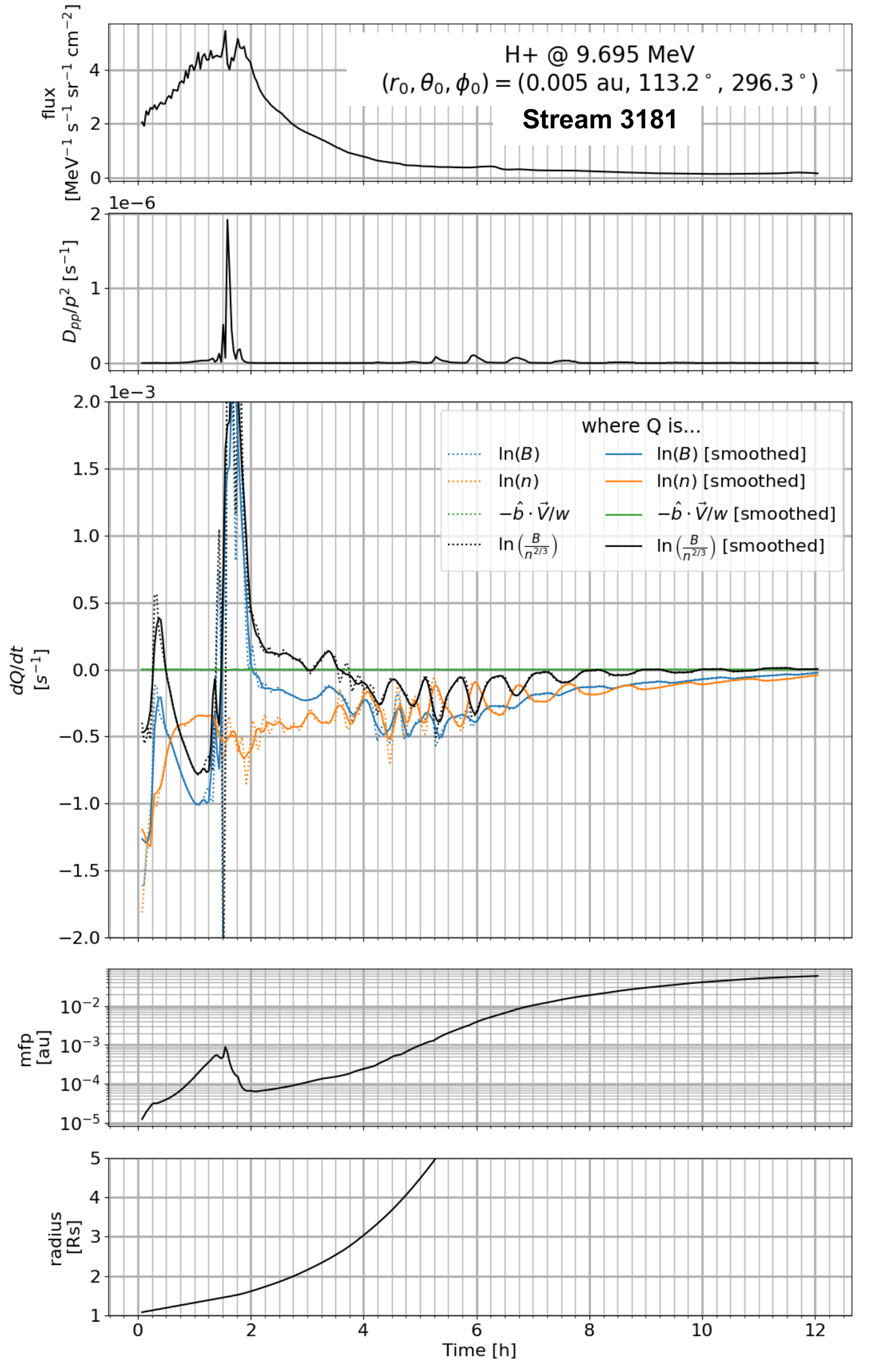}
\caption{The rate of change for quantities that drive acceleration in equation~(\ref{eq:focused_transport}) (third panel), the differential flux at 10~MeV/nuc (first panel),  and the radial distance from the Sun (bottom panel) of a node along streamline 3181 (shown in Figure \ref{fig:DASL}).  Also shown is the rate associated with stochastic acceleration (second panel) and the scattering mean free path used in the simulation (fourth panel). The dominant term driving acceleration is the rate of change of the magnetic field flux. As the node moves out approximately 3--8~hr into the simulation and traverses  a region ${\sim}2$--3.8~$R_{\odot}$, we observe several large reductions and then increases in the field strength associated with a current near the QSL boundary.  }
\label{fig:node3181}
\end{figure*}

\begin{figure*}[htbp!]
\centering
\includegraphics[width = 0.7\textwidth]{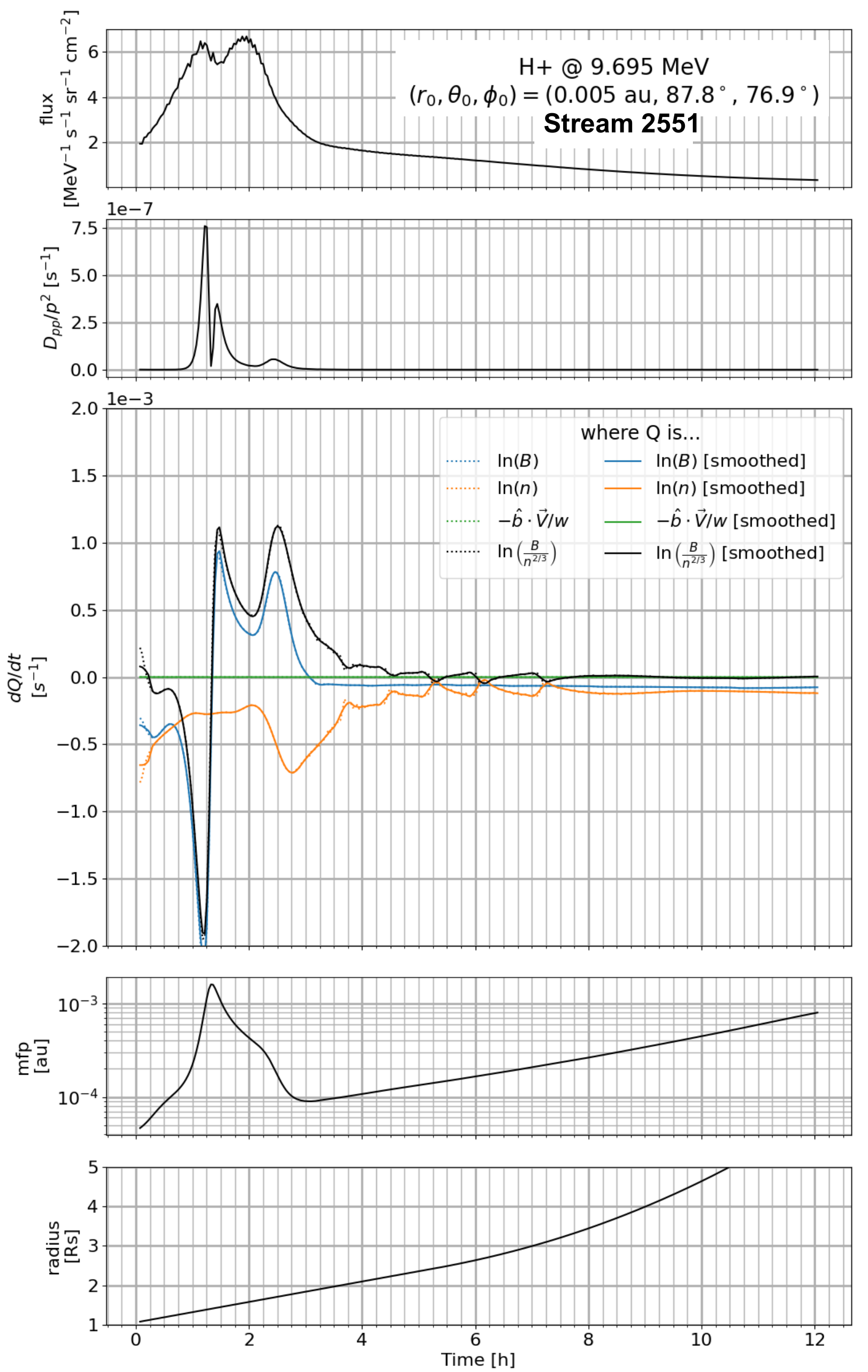}
\caption{The rate of change for quantities that drive acceleration in equation~(\ref{eq:focused_transport}) in a format similar to Figure~\ref{fig:node3181} for streamline 2551 (shown in Figure \ref{fig:DASL}).  Streamline 2551 moves through a QSL well way from the current sheet boundary.  }
\label{fig:node2551}
\end{figure*}

Figures~\ref{fig:node3181} and \ref{fig:node2551} show the the rate of change for quantities that drive acceleration in equation~(\ref{eq:focused_transport}) along these two streamlines.  The convective derivative of a given MHD quantity is evaluated  by differencing values  between points in time along the node history. Figures~\ref{fig:node3181} and \ref{fig:node2551} reveal that the dominant particle acceleration occurs where we observe a strong gradient in the magnetic field strength that is not accompanied by a strong expansion or compression in the density. In fact, as a node travels into the acceleration region, it first experiences a strong reduction in the magnetic field flux followed by an equivalent enhancement  in the field flux. These regions occur naturally where separatrix layers create reductions in the magnetic field magnitude, as illustrated in pseudo-streamer in Figure~\ref{fig:illustration}.

\section{Analytic Model of Particle Acceleration at a Separatrix Layer}
\label{sec:analytic}

In \S \ref{sec:qslEnergeticParticlesEffects} we began with a Focused Transport Equation, and considered acceleration due to rapid changes in the magnetic field strength and density. We performed a Legendre Polynomial expansion of the distribution function to reduce the Focused Transport Equation into a Parker-like transport equation, given by equation~(\ref{eq:transport}), for the isotropic-part of the distribution function. In addition to the standard Parker transport terms, we find a new term $ -p^{-2} \partial/\partial p [ p^2 D_{pp} \partial f_0/\partial p]$ showing that second-order Fermi acceleration (due to magnetic pumping)  operates at the separatrix layer. 

Appendix \ref{sec:bothProcesses} develops an analytic solution to equation~(\ref{eq:transport}) at the separatrix layer. We highlight several important aspects of this solution. The separatrix layer shares one similarity with a compression region in that the acceleration is spatially localized. However, unlike the compression region -- where particles gain energy on each crossing through a first-order Fermi process -- the separatrix layer allows for both energy gains and losses. The outcome depends on the particle pitch angle and the location of individual scattering events within the layer. Consequently, the particle acceleration mechanism at the separatrix layer is both localized and fundamentally a second-order Fermi process.

The analytic model depends on four parameters: 
\begin{itemize}
\item The quantity $\Delta u/u$ characterizes the overall compression through the separatrix layer. Some level of compression often accompanies the separatrix layer due in part to modification of the open field expansion:
\item The quantity $D$ is a dimensionless term that characterizes the diffusion in momentum space. Specifically, we take
\begin{eqnarray}
\frac{D_{pp}}{p^2}  & = &  D_0 L \delta(z)  \\
D_0 & = & \frac{\tau_s}{15} \left[ \frac{d \ln (B/n^{2/3})}{dt}\right]^2  \\
       & = & \frac{D}{t_0} \left( \frac{p}{p_i}\right)^\alpha  
       \label{eq:d0}
\end{eqnarray}
where $\tau_s = \lambda_\parallel / v $ is the scattering time, $v$ is the particle speed, and $p_i$ is the injection momentum. 
\item The quantity $\alpha$ represents the power-law momentum dependence of the diffusion coefficient $D_0$ (see equation~\ref{eq:d0}). 
\item The quantity $E_\mathrm{esc}$ represents a characteristic escape energy, which is detailed in \S \ref{sec:escape}. 
\end{itemize}

\begin{figure}
\centering
\includegraphics [width=0.8\columnwidth]{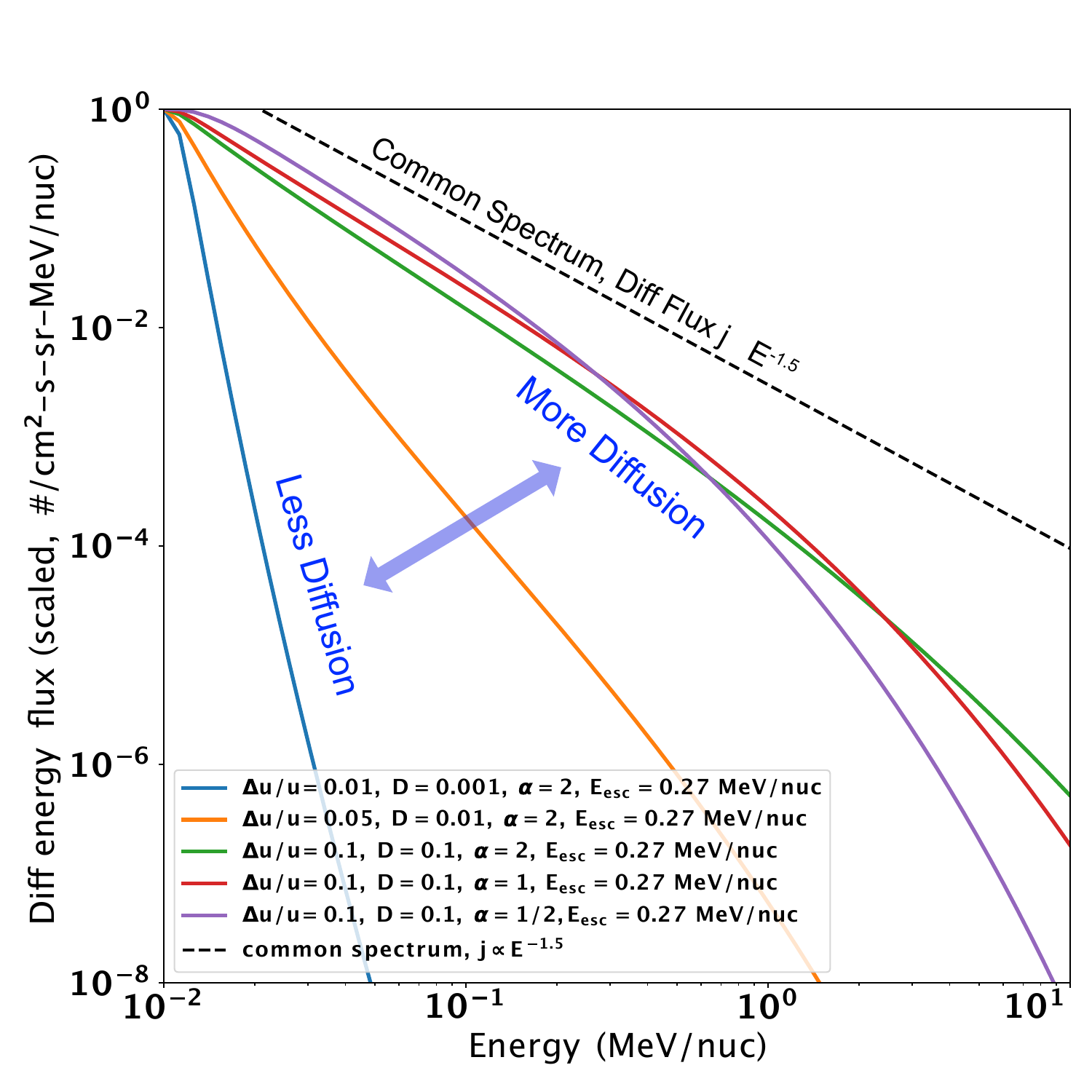}
\caption{Differential flux based on numerical integration of equation~(\ref{eq:hyz6}). The absolute scaling is derived  from the injection flux. Curves are shown as a function of energy for various levels of diffusive acceleration, momentum diffusion, and the scaling of diffusion with momentum as detailed in the appendix. High levels of diffusion converge to a common spectrum beneath the escape energy with differential energy flux proportional to $E^{-1.5}$. Low levels of diffusion result in a very soft spectrum. The diffusive power-law has a distribution function that scales as $p^{-\gamma}$. 
} \label{fig:diff3} 
\end{figure}

Results of the analytical model are shown in Figure \ref{fig:diff3} for an array of levels of $\Delta u/u$, $D$, and $\alpha$. In each case, we take $E_\mathrm{esc} = 0.27$ MeV/nuc. Higher rates of diffusion result in a spectrum close to the so called ``common spectrum'' with differential flux $\propto E^{-1.5}$ or speed $\propto v^{-5}$, as discussed by \cite{Fisk:2008}, \cite{Hill:2009}, and \cite{Dayeh:2009}. The fact that the acceleration mechanism recovers the common spectrum for large rates of diffusion is not accidental. This aspect of the acceleration is discussed and generalized in \S \ref{sec:random}. 

\begin{figure}
\centering
\includegraphics [width=0.8\columnwidth]{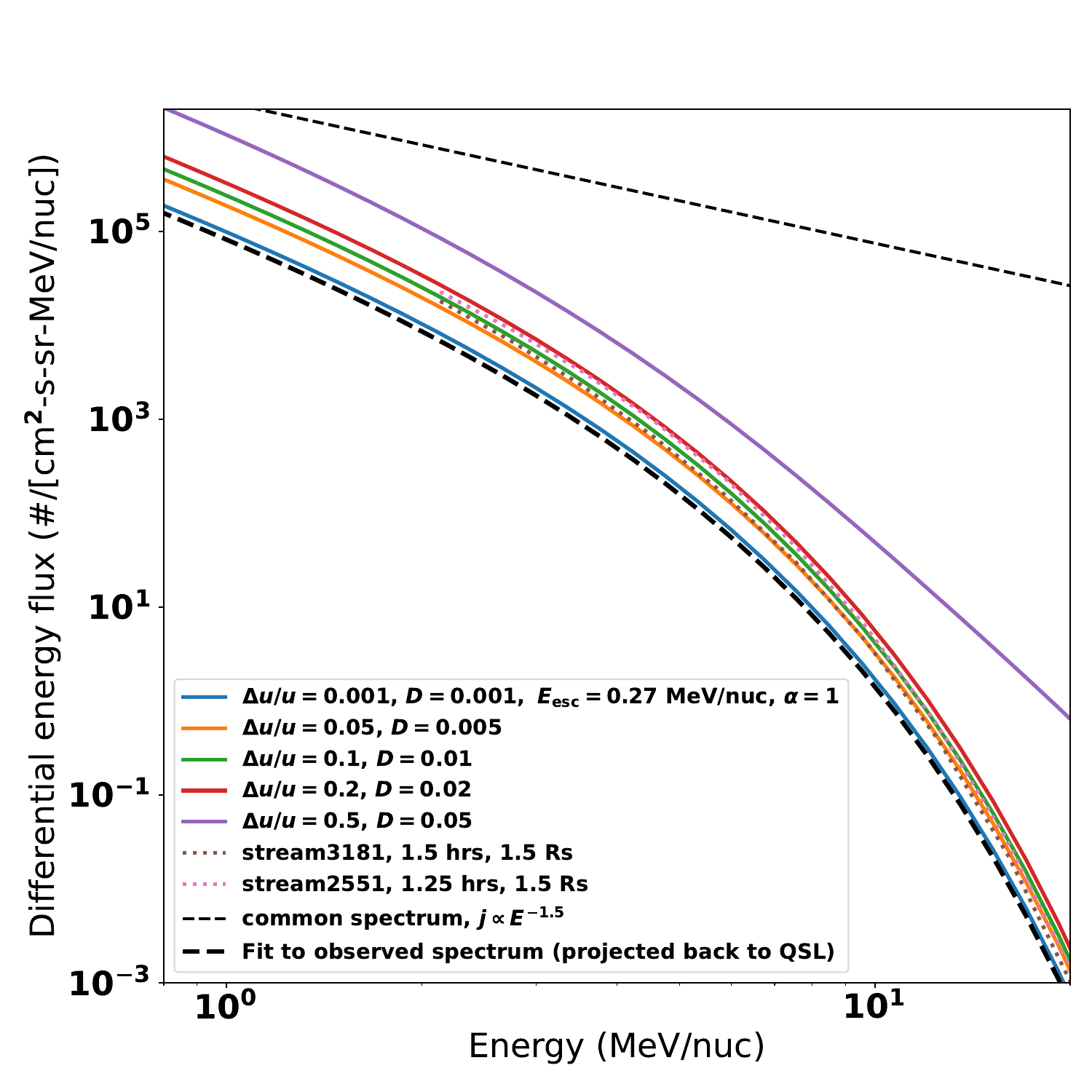}
\caption{Differential flux integrated across energy over the injection population in equation~(\ref{eq:hyz6}). Results are compared to  the STAT relaxation run. Curves are shown as a function of energy for various levels of diffusive acceleration, momentum diffusion, and the scaling of diffusion with momentum as detailed in the appendix. The seed population used in STAT is shown as the thick dashed curve, and solid lines show the energetic particle population for varying levels of second-order momentum diffusion. Dotted curves show the energetic particle fluxes from STAT for streams 3181 and 2551. Best agreement with stream 3181 is found for $D\sim 0.005$, and for stream 2551 with a higher level of momentum diffusion $D\sim 0.01$ - $0.02$. The thin dashed curve shows the hard common spectrum. 
} \label{fig:diff4} 
\end{figure}

It is important that we connect the analytic model of particle acceleration at the separatrix layer to the numerical model of the Focused Transport Equation. The numerical model captures the detailed behavior of the magnetic field, the separatrix layer, and includes changes in the scattering mean free path as function of distance from the Sun based on the strength of the magnetic field. In contrast, the analytical model represents an idealized and completely localized description of the acceleration process. The solution is, in a sense, a superposition of low-level compression and localized second-order Fermi acceleration at the separatrix layer. The second-order Fermi acceleration is the dominant process in the simulations performed. 

A key difference between the numerical model and the analytical model developed in Appendix \S \ref{sec:bothProcesses} is that acceleration in the analytic model proceeds from a specific injection momentum $p_i$ whereas the numerical model considers the seed spectrum and allows injection across all momenta solved for within the model. In other words, in order to compare the analytic model to the numerical model, we must integrate the injection across all momenta.  

The diffusion coefficient is determined by the gradients within the separatrix layer. We have applied the model with $\alpha = 1$, and an escape energy of 0.27 MeV/nuc. Tractable analytical solutions are restricted to positive values of $\alpha > 1/2$, and accurate computations are fairly straightforward for  $\alpha \ge 1$. The numerical solutions of the EPREM model does not have any such restrictions. Our simplified approach to comparing the analytic model with EPREM is to determine the diffusion coefficient at $\sim 1$ MeV/nuc, and  to then use that single diffusion coefficient over the entire energy range. A more complex analytical model can be implemented by adjusting the diffusion coefficient used over the range of integration. We have found the simplified model is a fairly accurate approximation to the more detailed analytical solution. 
 
We use a parallel scattering mean free path of $\sim 0.001$ au, a rate of change of magnetic field of $d \ln B/dt \sim 0.002 $ s$^{-1}$, a separatrix scale-size of $L = 0.2$ $R_\odot$ and a flow speed of 70 km/s into the separatrix layer. With these parameters, we find a diffusion coefficient $D_0 = 2.9\times 10^{-6}$ s$^{-1}$ at 1 MeV/nuc, and $D = 0.0058$. The analytical model with $D = 0.005$ agrees well with the EPREM simulation of streamline 3181 (see Figure 7). The reduction in field strength is readily computed based on the gradient. With the density gradient at 5\% of the of the field gradient, we find $\Delta u/u = 0.05$.

Figure \ref{fig:diff4} shows the comparison between the spectra from individual streamlines (3181 and 2551) and the analytical model. We find best agreement for $D$ between 0.005  and 0.02, which seems reasonable given calculation from the field gradient.  The two streamlines (3181 and 2551) are shown on the global configuration in Figure \ref{fig:DASL}. Streamline 3181 is near a current sheet, whereas streamline 2551 is close to a pseudo-streamer. The fluxes observed from streamline 2551 are larger than those from 3181, indicating high rates of acceleration. Comparing Figures \ref{fig:node2551} and \ref{fig:node3181}, we find that streamline 2551 shows multiple structures over a larger scalesize of more than 0.5 R$_s$. In contrast, the streamline 3181 is a compact structure associated with lower fluxes. The reduction in the diffusion scale $D$ for 3181 is due in large part to the smaller scalesize of the separatrix layer. 
 
\section{Connection with Random Stochastic Processes}
\label{sec:random}

The treatment of  separatrix layers results in a simple interpretation for the acceleration mechanism at work, while also suggesting a broader connection to the superposition of stochastic processes \cite[]{Schwadron:2010h}. Notably, the magnetic pump is in the class of superposition solutions, \cite{fisk:2010}. This is important here as it connects particle acceleration from stochastic fluctuations in the magnetic field magnitude with the superposition theory. 

In Appendix \ref{sec:DSA}, we find that the diffusive acceleration process proceeds along an acceleration characteristic, with the distribution function varying with an inverse exponential in time for energies below the escape energy (see equation~(\ref{eq:exptime})).  The fixed characteristic for acceleration depends on the presence of a first-order Fermi process. In contrast, the separatrix layers create variations in both strength of the magnetic field and in the density that result in second-order Fermi acceleration  (see equation~(\ref{eq:transport})). The acceleration in this case does not occur strictly along acceleration characteristics. Instead, there is an array of characteristics that are populated as energy diffusion broadens the energy distribution as a function of time. Fundamentally, the dependence on an inverse exponential in time remains (\ref{eq:exptime2}). But the final solution becomes much more complex since many different inverse exponential terms contribute, depending on the relative time of particle injection and the broadening through energy diffusion. 

The presence of inverse exponential time-dependence suggests a deeper connection with stochastic processes. Appendix \ref{sec:waiting} develops this connection in detail, beginning with the concept of waiting time distributions from statistical theory, and leading to the following expression for the distribution function, $f$, for particle speeds greater than the core or thermal speed of the distribution ($v > c_s$), given a distribution density $n$:
\begin{eqnarray}
f(v > c_s) & = & n P_s(v)/(4 \pi v^2) \nonumber \\
 & = & \left( \frac{n}{2\pi v^2}  \right) \frac{ \zeta_0}{\left( v \zeta_0 + 1\right)^3}\,. 
 \label{eq:kappa}
\end{eqnarray}

An interesting result that follows from this distribution is the $v^{-5}$ distribution at high particle speeds. We note that this distribution is similar to that found in \S \ref{sec:analytic} in the limit of strong diffusion. While there is no one acceleration rate at all energies, there is a rate of diffusion in momentum space, which controls the overall rate of acceleration.  Over short periods of time compared to the overall acceleration time, the change in speed varies approximately linearly with time. The tail of the accelerated population reveals the superposition of states in the  distribution. The superposition of exponential states of particle speed is inevitable (see equation~(\ref{eq:P5})) provided that particles are not subject to rapid acceleration by a specific first-order mechanism, or that states are not overtaken, as occurs in shocks, or in non-linear cases when the acceleration decreases with particle speed. In such exceptions, the buildup of states collapses to a specific velocity and  behaves in a manner similar to a soliton solution. 

The diffusive process provides a build-up in particle speed for the distribution through multiple encounters of fluctuations in the magnetic field magnitude and the density of the plasma. Each of these fluctuations behaves as an individual state with a specific acceleration and approximately exponential dependence on particle speed. Since diffusion is inherently random, the entropy of states must be maximized. Therefore, in the limit of strong diffusion, the superposed distribution must follow the kappa function (\ref{eq:kappa}). 

The agreement between the kappa function and the more detailed analytic treatment in  \S \ref{sec:bothProcesses} and shown in \S \ref{sec:analytic} is not accidental. However, the diffusive and first-order processes dealt with previously (\S \ref{sec:analytic}) require an injection speed where particles are capable of moving upstream in the plasma. The superposition of states does not require an injection speed. The difference is that acceleration acts in the frame of the solar wind, and operates as a continuum of stochastic processes or states. This occurs provided that there is a turbulent cascade from the sites of reconnection that drive the fluctuations in the plasma. 

The main parameters that enter the estimation of the stochastic distribution are the density and the mean inverse speed $\zeta_0$. Near the QSL particles gain energy through momentum diffusion. It is mesoscale and small-scale structures in the solar wind that provide acceleration. The source of energy for these structures is reconnection within the QSL. The QSL structures cascade down in size as they interact with the surrounding solar wind, driving energy from larger to smaller scales.  The rate of momentum diffusion $D_{pp}^\ell$ in these smaller structures is derived based on their sizescale $\ell$ and the sizescale $L$ of larger region near the QSL,
\begin{eqnarray}
    D_{pp}^\ell = \frac{\ell}{L} D_{pp}\,.
    \label{eq:rate}
\end{eqnarray}
The mean inverse speed is tied to acceleration in the mesoscale and small-scale structures,
\begin{eqnarray}
    \frac{1}{\zeta_0} = \sqrt{ \frac{D_{pp}^\ell(c_s) }{m^2} \frac{\ell}{c_s} }
    \label{eq:zeta0}
\end{eqnarray}
where the momentum diffusion rate $D_{pp}^\ell (c_s )  $ is evaluated at the sound speed $c_s$.

\section{Seed Population Fluxes from the QSL region}
\label{sec:qsFluxes} 

Another aspect of particle acceleration in QSLs is that it may account for the generation of seed populations.
Within the QSL, magnetic reconnection occurs frequently between closed magnetic structures and the open field lines that guide the solar wind. Particle acceleration near the QSL requires an injection energy $\sim 0.01$~MeV/nuc such that particles can move upstream and then, through multiple encounters of particles moving upstream and downstream, interact with the QSL repeatedly. 

The structures created by interactions near the QSL cascade down in size as they interact with the surrounding solar wind, driving energy from larger to smaller scales. At energies below the injection energy, particles interact with and are accelerated by these meso-scale structures and small-scale structures in the solar wind (\S~\ref{sec:random}), resulting in the distribution function given by equation~(\ref{eq:kappa}). The quantity $\zeta_0$ in that equation represents an inverse mean speed characteristic of superposed distributions. The superposition sums weighted exponential probability distributions, $\exp( - \zeta v)$, each with different exponential roll-overs at inverse speed $\zeta$.   The array of these weighted probability distributions are characteristic as states in a system, and the weights are determined from the maximization of Boltzmann entropy  \cite[]{Schwadron:2010h} with the constraint of mean inverse speed $\zeta_0$.  The acceleration of particles associated with the states in the system is driven by momentum diffusion, with a rate that is scaled based on the size of the meso- or small-scale structure versus the size of the larger-scale QSL (cf. equation~(\ref{eq:rate})). The inverse mean speed, in turn, is related to the timescale of interaction multiplied by the diffusion rate evaluated near the sound speed of the plasma (cf. equation~(\ref{eq:zeta0})). The distribution function in equation~(\ref{eq:kappa}) thus represents the population of particles injected into higher energy acceleration near the QSL. 

\begin{figure*}[th!]
\centering
\includegraphics[width = 0.8\textwidth]{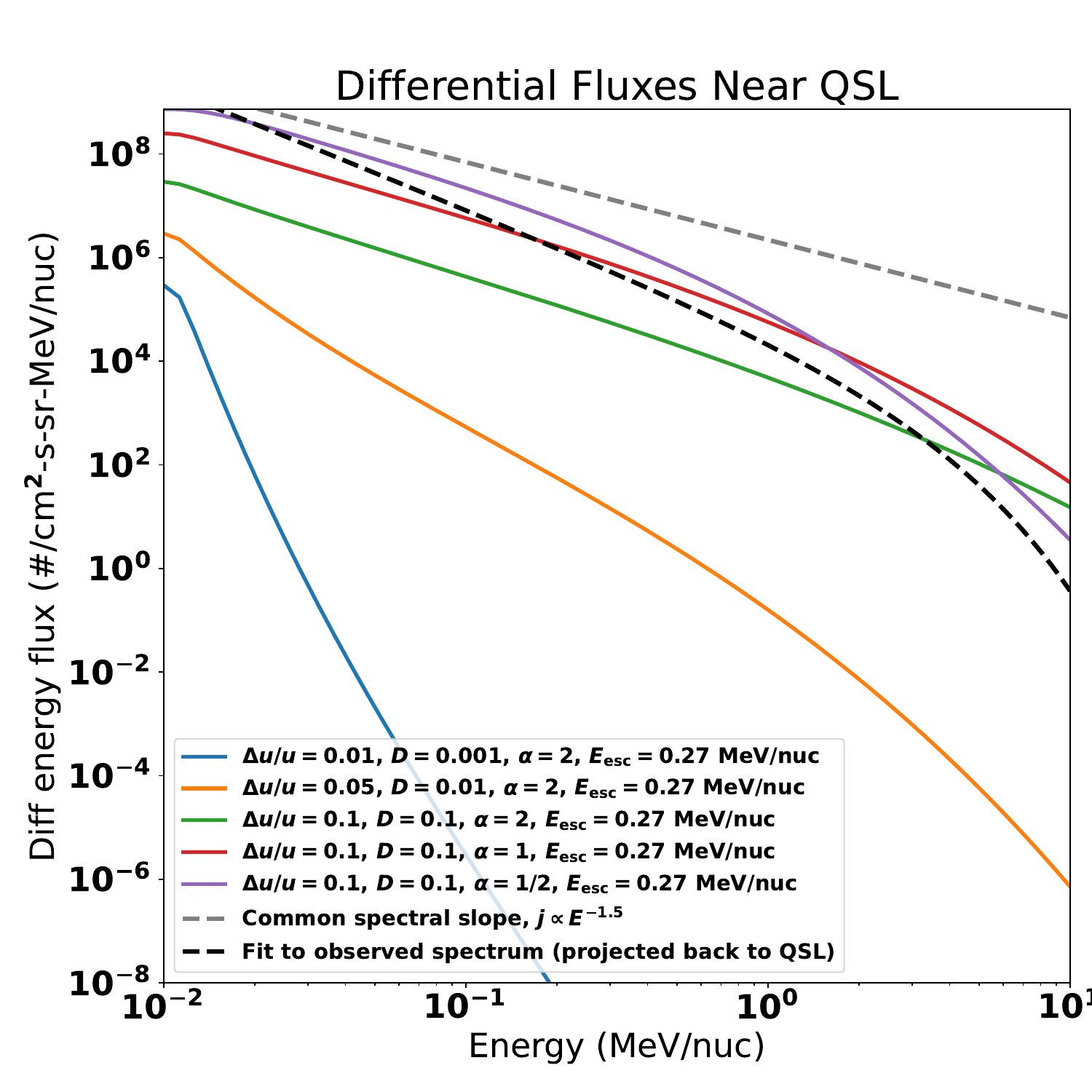}
\caption{
Differential fluxes derived from numerical integration of equation~(\ref{eq:hyz6}). Resulting distributions are the result of particle acceleration near QSLs. Curves are shown as a function of energy for various levels of diffusive acceleration, momentum diffusion, and the scaling of diffusion with momentum as detailed in the appendix. The amplitude of the differential fluxes is tied to the injection of particles accelerated from meso-scale and small-scale structures in the solar wind. High levels of diffusion converge to a common spectrum beneath the escape energy with differential energy flux proportional to $E^{-1.5}$ (see top grey dashed line). Low levels of diffusion result in a very soft spectrum, with relatively low differential fluxes. Also shown is the seed spectrum used for modeling near the QSL at 3 $R_{\odot}$. This seed spectrum is fit from PSP observations at 0.8~au and projected back to the QSL near the Sun.
\label{fig:diff4Seed}}
\end{figure*}

Figure~\ref{fig:diff4Seed} shows calculated distributions above the injection energy. The amplitudes of the differential fluxes are determined by the injected particles accelerated by meso-scale (and small-scale) structures, and the subsequent acceleration in the large-scale QSL at energies greater than the injection energy. We have taken a plasma density $n$ projected to the QSL near 3~$R_{\odot}$ from the location of PSP (0.8~au) using a density at PSP of roughly 5~cm$^{-3}$.  The transition from small-scale to meso-scale structure is at the size of $\ell \approx 5$~Mm based on observations \cite[]{Viall:2021}.  This transition occurs between the solar wind kinetic scales and the meso-scale structures that begin to form in the spatial range of $\ell > $ 5~Mm. 

\begin{figure*}[th!]
    \centering
    \includegraphics[width = 0.8\textwidth]{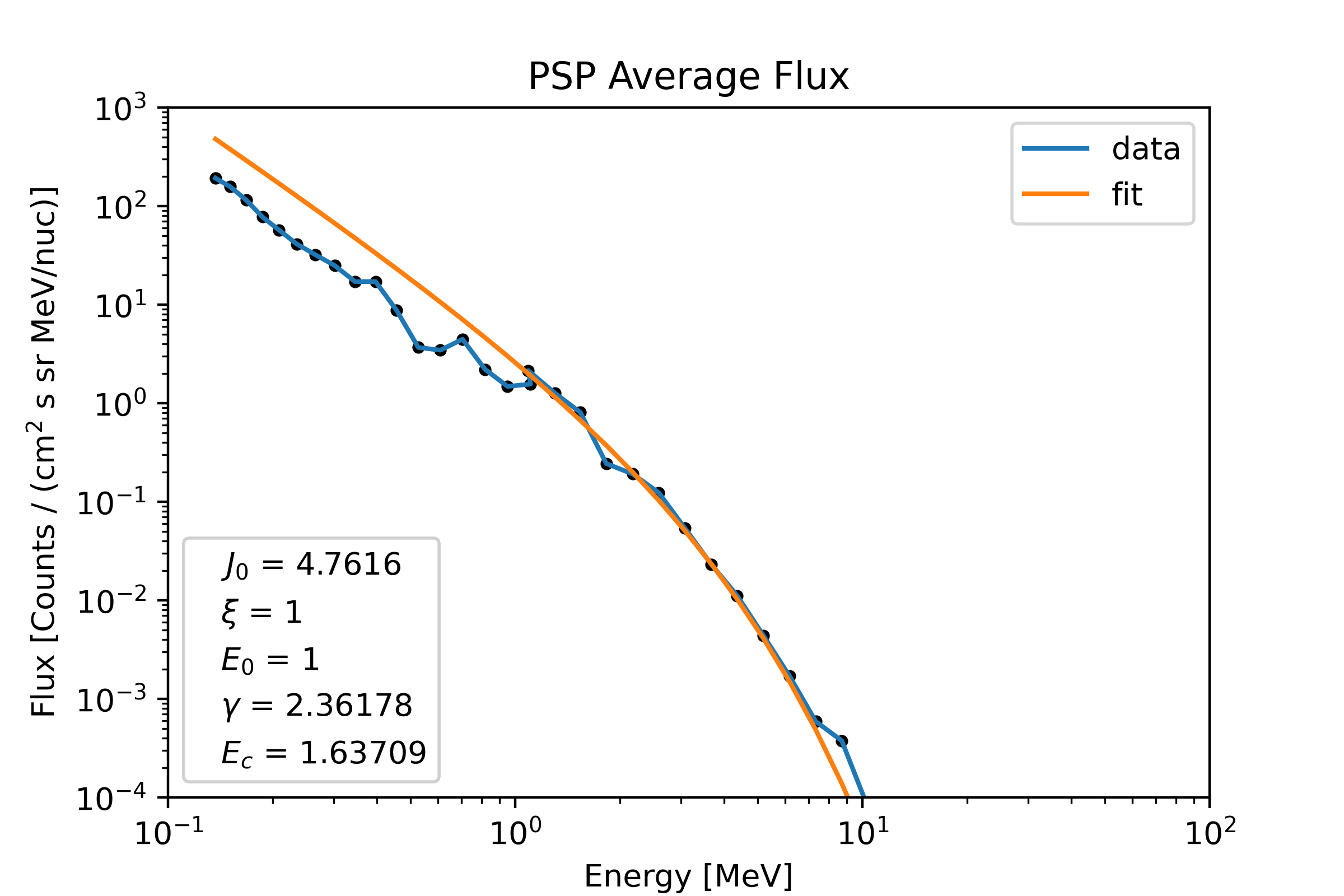}
    \caption{Observed proton fluxes from PSP/EPI-Lo and PSP/EPI-Hi (black dots with blue line) and fit to the analytic form of the seed-spectrum equation used in EPREM numerical solutions. The legend in the lower left gives the values of equation parameters.}
    \label{fig:seed-spectrum-psp-fit}
\end{figure*}

Also shown in Figure~\ref{fig:diff4Seed} is the seed spectrum (black dashed curve) used for event modeling. This seed spectrum was fit to PSP EPI-Lo and EPI-Hi average proton fluxes during the pre-event period 00:00 -- 10:00 UTC on November 29, 2020, using the following form, 
\begin{eqnarray}
    J_\mathrm{seed}(E, r = r_0)  =  J_0 \left( \frac{E}{E_0} \right)^{-\gamma} \exp \left( - \frac{E}{E_c}\right)
    \label{eqn:seed-spectrum-analytic}
\end{eqnarray}
and the seed spectrum is projected to the QSL near 3~$R_{\odot}$ based on a $1/r^\beta$ radial dependence. The fit to observations at PSP, which was located approximately 0.8~au from the Sun, yields $J_0 \approx 4.8$ particles/(cm$^2$-s-sr-MeV), $\gamma \approx 2.4$, $E_\mathrm{c} \approx 1.6$ MeV, and we have taken a standard square-radial distance scaling, $\beta = 2$. The reference energy, $E_{0}$, is set at 1 MeV. Figure \ref{fig:seed-spectrum-psp-fit} shows the observed flux values along with a non-linear least squares fit of equation~(\ref{eqn:seed-spectrum-analytic}) to the observations. Curve fitting was performed using \texttt{scipy.optimize.curve\_fit} from SciPy version 1.6 \citep{2020SciPy-NMeth}. In order to use the fit spectrum as the seed spectrum for EPREM within STAT, we scaled the amplitude to a radial distance of 1.0~au via the same $1/r^\beta$ power law, giving a value of 3.1 particles/(cm$^2$-s-sr-MeV). This value, when scaled to the radial distance of the QSL, provides the amplitude of the seed spectrum in Figure~\ref{fig:diff4Seed}.

The rates of diffusion are characterized based on the dimensionless quantity $L D_0/u$ defined as
\begin{equation}
\frac{L D_0}{u} = \frac{\tau_s}{15}\frac{L}{u} \left[ \frac{d \ln (B/n^{2/3})}{dt}\right]^2 = D \left( \frac{p}{p_\mathrm{inject}}\right)^\alpha\,,
\end{equation}
\noindent where $L$ is the characteristic size of the diffusive acceleration region and $u$ is the wind speed through the reconnection exhaust. The quantity $p_\mathrm{inject}$ represents the injection momentum, $\alpha$ provides the scaling with particle momentum, and $D$ is a dimensionless quantity for the amplitude of momentum diffusion. Similarly,  the compression or shock is characterized by the relative change in plasma flow velocity, $\Delta u/u$, across the shock or compression. Results of solutions indicate that strong diffusion yielding a common characteristic spectrum with differential energy flux $j \propto E^{-1.5}$ occurs in the regime where $L D_0/u > 0.07$, whereas diffusion rates $L D_0/u < 0.005$ yield very soft spectra. 

The results shown in \S 4 indicate that relatively weak diffusion $L D_0/ u \sim 0.01$ from separatrix layers can account for the results observed from STAT when a pre-existing seed population is present. Answering the question of how this seed population is generated to begin with relies on additional acceleration from a cascade of magnetic field energy released by separatrix layers. Generating this seed population requires rates of momentum diffusion that are nearly five to ten times that found in the STAT simulations. It is plausible that these sources exist since the coronal structure is highly complex and the structures simulated are on a large-scale compared to the hierarchy of structures in the corona. Further, the simulations show only conditions in an idealized setting with a relaxed coronal field. Transient structures are always present in the corona and will drive magnetic energy from the separatrix layers into the solar wind. 

 
\section{Discussion} 
\label{sec:composition}

The acceleration detailed in this manuscript is driven by magnetic field changes at separatrix layers. The reduction in field strength at QSLs precondition these sites to magnetic reconnection. After reconnection, the magnetic fields are reconfigured as they relax toward a potential field configuration.  \textcolor{black}{The reconfiguration of the magnetic field is similar to a flaring process.} The reconnection process also releases material previously contained within closed magnetic field structures, which are often rich in heavy ions and would store $^3$He enriched plasma. Therefore, the seed population produced by QSLs is expected to be rich in $^3$He and heavy ions. 

Pre-existing seed populations were hypothesized as the result of an array of flares or nanoflares ongoing at the Sun \cite[]{Parker:1988}. The presence of enhanced $^3$He throughout observed events provides direct evidence that flares contribute to energetic particle seed populations \cite[]{Mason86, Mason:2002, Reames:1999, Desai:2003}. The seed populations from flares are  compressively enhanced (pre-conditioning the seed populations), and then accelerated to higher energies as strong compressions and shocks form further out in the heliosphere. The steps that involve the release, compression and further acceleration support the unifying role of the CME, together with the production of seed populations from  flaring. 

The build-up of heavy ions on closed field structures was recently corroborated by direct observations from PSP \cite[]{Schwadron:2024}.
PSP/IS$\odot$IS at $\sim$0.2~au on 
March 2, 2022 observed an event in which there was rare coincidence between 
imaging and in situ measurements. During this event, PSP passed through structures on the flank of a 
streamer-blowout CME including an isolated flux tube in front of the CME, a turbulent sheath, and the CME itself.  
The isolated flux tube shows large streaming, hard spectra, and large Fe/O and He/H ratios, indicating flare sources. 
Energetic particle fluxes are most enhanced within the CME interval from suprathermal through energetic particle energies
($\sim$ keV to $>10$~MeV), indicating particle acceleration,
as well as confinement local to the closed magnetic structure. The flux-rope morphology of the CME helps to enable local modulation and trapping of energetic particles. The closed field structure of the CME acts to build up energetic particle populations. 

The CME observed by \cite{Schwadron:2024} is consistent with the the build-up of suprathermal and energetic particle seed populations with heavy ion enhancements on closed field structures. Within  QSL reconnection regions, we therefore expect the release of seed populations that have elevated levels of heavy ions.

\section{Conclusion} \label{sec:conclusion}

This paper provides an analytical treatment of particle acceleration near the Sun at quasi-separatrix layers.    The analytic treatment shows that a second-order acceleration process occurs as ions traverse the QSL regions where the magnetic field magnitude varies strongly in response to the complex coronal field configuration. 

\textcolor{black}{
The energy in magnetic field changes near the QSL constitutes fluctuations in the magnitude of the magnetic field. In the presence of plasma flow along the open field and through the QSL, magnetic pumping results in the acceleration of  SEPs. We have used the STAT  model to show the first global simulations of energetic particles accelerated from QSLs and above current sheets at the Sun, which reveals these mechanisms at work in the low corona.
}

Our results address a number of key aspects of particle acceleration in the low corona: 
\begin{itemize}
\item The presence of field-magnitude changes above QSLs provides a direct means for second-order  particle-acceleration, as shown in \S \ref{sec:qslEnergeticParticlesEffects}. The acceleration process can be approximated using a Parker-type transport equation with the addition of a second-order acceleration term involving diffusion in momentum space, where the diffusion coefficient scales with the square of fluctuations in $\ln B/n^{2/3}$  and the scattering time. 
\item An analytic solution that includes the acceleration resolves exponential roll-over solutions (roll-over $\sim 0.2 - 2$ MeV/nuc) that are commonly observed in seed populations of energetic particles. 
\item An important feature of these solutions is the emergence of low-energy power-laws that \emph{approach} but are still softer than the $j \propto E^{-1.5}$ spectrum often reported at low energies ($3 - 50$ keV/n) \cite[]{Gloeckler:2002a, fisk:2010}. 
\item Separatrix layer acceleration of suprathermal ions down to energies that approach the low-energy core plasma population emerges from a cascade-like process to generate small-scale irregularities in the solar wind from the large-scale features released by the QSL. The cascade process prevents an injection problem for the energetic particle acceleration, and allows particle acceleration to continue in the solar wind reference frame through successive interactions with small-scale irregularities in the magnetic field magnitude. 
\item The interaction of energetic particles with small-scale irregularities can be treated as the superposition of waiting-time distributions, where the waiting timescale multiplied by the characteristic acceleration rate yields a mean-speed tied to the rate of the acceleration processes acting within the cascade. The resulting superposed particle distribution has a kappa-like character down to the core of the solar wind and roughly matches the power-law at larger energies obtained from energy diffusion. 
\item The combination of acceleration  processes involving small-scale irregularities that cascade to small sizescales, and larger scale irregularities that drive acceleration directly from separatrix layers provides for significant fluxes of energetic particles that are similar to the fluxes observed in suprathermal proton populations. 
\item \textcolor{black}{The reconfiguration of the magnetic field involved in separatrix layer acceleration bears similarities to a flaring process.} The reconnection process releases material previously contained within closed magnetic field structures, which are typically rich in $^3$He and heavy ions. Both the connection to flaring and the release of previously stored plasma on closed magnetic structures indicates that seed populations accelerated in separatrix layers should be rich in $^3$He and heavy ions. 
\item We utilized the STAT model to validate the theoretical results described, where particle acceleration from the QSL regions was observed as the theory predicts.
\end{itemize}
\textcolor{black}{Thus, we provide analytical results and numerical simulations  suggesting that separatrix layers provide a source of magnetic irregularities on open field lines that drive the acceleration of energetic particles, and therefore provide a plausible source for energetic particle seed populations near the Sun. } 

The results provide  a novel explanation for broad-longitudinal SEP events generated by CMEs.  CMEs could disrupt the coronal field configuration, which would cause significant magnetic reconnection at QSLs. This reconnection would in turn enhance particle acceleration, and thereby generate larger numbers of suprathermal and energetic particles from the separatrix layers.  We plan to  use STAT to explore this mechanism in future publications.

\section*{Acknowledgements}
The authors would like to thank T.~Forbes for discussions, almost a decade before submission, that helped to formulate ideas used in the manuscript. The authors would like to thank T.~Lim and C.-M.~Um for checking the derivation of the transport equation in Appendix~A. 
Support for the project was provided by NASA's LWS-SC (grant no.\ 80NSSC22K0893) and LWS (grant no.\ 80NSSC19K0067) programs.


\bibliographystyle{aasjournal}

\appendix



\section{Acceleration at a 2D Structure} \label{app:acc2d}

Our approach to understanding the influence of structure on the evolution of distribution functions in the outflows of the separatrix layer utilizes the simplified transport equation~(\ref{eq:transport}) as a starting point. 
We define the coefficient C \cite[the Compting-Getting coefficient, ][]{Forman70, Rygg:1971} for the distribution function, and allow for changes in the power-law,
\begin{eqnarray}
p\frac{\partial f_0}{\partial p} = - 3C f_0\,.
\end{eqnarray}
We recast the transport equation in a conservative form: 
\begin{eqnarray}
\frac{\partial f_0}{\partial t} + \nabla \cdot \left(\mathbf{u} f_0 \right) - \nabla\cdot \left( \bar{\bar{\kappa}} \cdot \nabla f_0 \right)  
 +(C - 1)\nabla\cdot \mathbf{u} f_0
 -  \frac{1}{p^2} \frac{\partial }{\partial p} \left( p^2 D_{pp} \frac{\partial f_0}{\partial p} \right)  = 0 \,.
\label{eq:transport2}
\end{eqnarray}
Here, we take $z$ as a linear coordinate along a magnetic field line, and the $z=0$ position  where the particles are fed into acceleration. Similarly, both the flow divergence ($ \nabla \cdot \mathbf{u}$) and the momentum diffusion term ( $\frac{1}{p^2} \frac{\partial }{\partial p} \left( p^2 D_{pp} \frac{\partial f_0}{\partial p}\right)$) are taken as localized acceleration at $z=0$:
\begin{eqnarray}
\nabla \cdot \mathbf{u} & = & - \Delta u \; \delta (z)  = - (d \ln n/dt ) L \delta (z) \\
D_{pp}/p^2 & = &  D_0  L \; \delta(z),
\end{eqnarray}
where $L$ is the characteristic size of the acceleration region, $\Delta u$ is the speed reduction across the interface, and
 \begin{eqnarray}
D_0  =  \frac{\tau_s}{15} \left[\frac{d \ln (B/n^{2/3})}{dt} \right]^2.
\end{eqnarray}

We take what has become a standard approach to diffusive acceleration, by first integrating the distribution upstream of the acceleration region, where convection balances the back diffusion of accelerated ions: 
\begin{eqnarray}
u_{su} f_0 - \kappa_\parallel \frac{\partial f_0} { \partial z} = 0, 
\end{eqnarray}
where $u_{su}$ is the flow speed parallel to the magnetic field upstream from the discontinuity. The upstream solution readily follows
\begin{eqnarray}
f_0 \propto f_a \exp(u_s z/\kappa_\parallel)
\end{eqnarray}
and downstream, $f_0 \approx  f_a$. 
We have denoted $f_a$ as the isotropic part of the distribution function associated with accelerated particles at the position of localized acceleration. Integrating across a small region local to the acceleration region, we recover the following
\begin{eqnarray}
u_{su} f_a + \frac{\Delta u}{3} p\frac{\partial f_a}{\partial p}      -  \frac{1}{p^2} \frac{\partial }{\partial p} \left( p^4 D_{0} L \frac{\partial f_a}{\partial p} \right) = 0
\label{eq:steadyState}
\end{eqnarray}
Note that the acceleration process involves diffusion, and as the scattering mean free path approaches the scale of the acceleration region, particles experience an increased probability of escaping the acceleration process. We therefore include a loss term associated with escaping particles:
\begin{eqnarray}
u_{su} f_a + \frac{\Delta u}{3} p\frac{\partial f_a}{\partial p}      -  \frac{1}{p^2} \frac{\partial }{\partial p} \left( p^4 D_{0} L \frac{\partial f_a}{\partial p} \right) = - \frac{f_a}{\tau_\mathrm{esc}}
\label{eq:steadyState2}
\end{eqnarray}
where the loss-timescale is energy dependent, $\tau_\mathrm{esc} = \tau_\mathrm{esc}(E)$. 

We proceed by first treating the problem of diffusive acceleration as a superposition of acceleration processes (\S \ref{sec:DSA}), and then proceed to derive the more general solutions that include both diffusive shock acceleration and energy diffusion from the QSL layer (\S \ref{sec:bothProcesses}). 

\subsection{Diffusive Shock Acceleration as an Exponential Process}
\label{sec:DSA}

We develop an approach that is consistent with a well-known  diffusive acceleration solution, while also providing a methodology that allows the problem to be compared to other treatments for the superposition of stochastic processes \cite[]{Schwadron:2010h}. Since we neglect diffusion, equation~(\ref{eq:steadyState2}) becomes,
\begin{eqnarray}
\frac{\partial f_a}{\partial t} + \frac{u_{su}}{L} f_a + \frac{\Delta u}{3L} p\frac{\partial f_a}{\partial p}      -   = - \frac{f_a}{\tau_\mathrm{esc}}
\label{eq:time_dep_DSA}
\end{eqnarray} 
By retaining time-dependence, we gain insight into the acceleration process that is useful both in understanding the relationship between diffusive shock acceleration  and superposed stochastic processes, and in deriving the solution that includes energy diffusion. In introducing time-dependence, we also must include a spatial scale $L$ over which the diffusion acceleration process proceeds. This spatial scale is related to the diffusion coefficient and the upstream solar wind speed, as detailed by \cite{Schwadron:2015SEP}.

It is convenient to introduce the following variables:
\begin{eqnarray}
t_0 & = & L/u_{su}, \\
s & = & t/t_0, \\
q & = & p/p_i,
\end{eqnarray}
where $p_i$ is the injection momentum. The dimensionless form of equation~(\ref{eq:time_dep_DSA}) is,
\begin{eqnarray}
\frac{\partial f_a}{\partial s} + f_a 
\left(1 + \frac{t_0}{\tau_\mathrm{esc}}\right)  + \frac{1}{\gamma} \frac{\partial f_a}{\partial \ln q}       = 0.
\label{eq:nodim_DSA}
\end{eqnarray}
where $\gamma = 3r_c /(r_c - 1) = 3 u_{su} /( \Delta u)$ and $r_c = u_{su}/u_{sd}$ is the compression ratio.

An important feature of equation~(\ref{eq:nodim_DSA}) is that the acceleration proceeds along a characteristic:
\begin{eqnarray}
q(s, s_i) = \exp[\Delta u (s-s_i)/ (3 u_{su}) ] .
\label{eq:char}
\end{eqnarray}
where $s_i$ is the time of particle injection. 
Taking $f_a = g \exp[-(s-s_i)(1+t_0/\tau_\mathrm{esc})]$ provides a simplification for treating the particle loss.  We neglect momentum gradients of the escape timescale, \textcolor{black}{which requires that
\begin{eqnarray}
\frac{\partial}{\partial \ln p}\left( \frac{t_0}{\tau_\mathrm{esc}}  \right)  \ll 1 .
\label{eq:condition}
\end{eqnarray}
In this paper we take the escape rate ($t_0/\tau_\mathrm{esc}$) to scale linearly with momentum or as a power weaker than linear. Equation (\ref{eq:condition}) therefore implies that $t_0/\tau_\mathrm{esc} \ll 1$ or equivalently that $E \ll E_\mathrm{esc}$. The solution is therefore a good approximation of the distribution function for energies below the escape energy. Above the escape energy, the distribution falls off more rapidly with energy, and our approximation provides a simplified form of the break. A more accurate numerical solution above the break energy shows an even more rapid decrease in the distribution function with energy. However, since the distribution drops more quickly in this regime, a more accurate approximation of the  decay in the  distribution function provides limited value. 
}

Taking $g = g(s,q)$  yields the following, 
\begin{eqnarray}
\frac{\partial g}{\partial s} = 0
\label{eq:g}
\end{eqnarray}

\newcommand{\tesc}{\tau_\mathrm{esc}}

Following the characteristic implies that 
\begin{eqnarray}
f_a = f_i \exp(-[s-s_i] [1+t_0/\tesc])
\label{eq:exptime}
\end{eqnarray}
where $f_i$ is the distribution function at the injection energy. We can solve for $(s-s_i)$ from the characteristic, 
\begin{eqnarray}
s - s_i  = \gamma \ln q
\end{eqnarray}
and 
\begin{eqnarray}
f_a  & = & f_i \exp\left( - \gamma[1 +t_0/\tesc]  \ln q  \right) \\
 & = &  f_i (p/p_i)^{-\gamma[1+ t_0/\tesc]}
\end{eqnarray}
Thus, for no escape,  we recover the standard diffusive shock acceleration power-law $f_a \propto p^{-\gamma}$.  The more general case that includes escape leads to broken power-law distributions as discussed by \cite{Schwadron:2015SEP}.
The result was derived using a model with time-dependent acceleration. This approach provides a deeper connection with stochastic processes, as detailed in \S\ref{sec:random}. 

\subsection{Solution Including Energy Diffusion and Diffusive Shock Acceleration}
\label{sec:bothProcesses}

We now take up the solution that incorporates both diffusive shock acceleration and energy diffusion. As in \S \ref{sec:DSA},  we find the steady-state limit, and time-dependence is considered as a means to simplify deriving analytic solutions. Therefore, the starting point is the following differential equation,
\begin{eqnarray}
\frac{\partial f_a}{\partial t} + \frac{u_{su}}{L} f_a + \frac{\Delta u}{3L} p\frac{\partial f_a}{\partial p}      -  \frac{1}{p^2} \frac{\partial }{\partial p} \left( p^4 D_{0} \frac{\partial f_a}{\partial p} \right) =  - \frac{f_a}{\tau_\mathrm{esc}} 
\label{eq:time_dep}
\end{eqnarray} 
Our initial approach is similar to the treatment of diffusive shock acceleration. 
We take 
\begin{eqnarray}
f_a = g \exp(-[s-s_i][1+t_0/\tesc]),
\label{eq:exptime2}
\end{eqnarray}
 where $g = g(s, q)$, and  neglect momentum gradients  (as in \S\ref{sec:DSA}) of the escape timescale since these gradients are far smaller than the momentum gradient in the distribution function \textcolor{black}{for energies less than the escape energy}. This substitution results in the following,  
\begin{eqnarray}
\frac{\partial g}{\partial s} 
+ \frac{1}{\gamma} \frac{\partial g}{\partial \ln q} 
 -  t_0 \frac{1}{q^2} \frac{\partial }{\partial q} \left( q^4 D_{0} \frac{\partial g}{\partial q} \right) = 0
\label{eq:g2}
\end{eqnarray}

\newcommand{\spr}{s^\prime}
\newcommand{\qpr}{q^\prime}

We now introduce generalized coordinates $(\spr,\qpr)$ that allow for acceleration along a characteristic, and diffusion across characteristics:
\begin{eqnarray}
\spr & = & s \\
\qpr & = &  q \exp(- s/\gamma)
\end{eqnarray}
In these coordinates, equation~(\ref{eq:g2}) becomes, 
\begin{eqnarray}
\frac{\partial g}{\partial \spr} 
 -  t_0 \frac{1}{(\qpr)^2} \frac{\partial }{\partial \qpr} \left[ (\qpr)^4 D_{0} \frac{\partial g}{\partial \qpr} \right] = 0
\label{eq:g3}
\end{eqnarray}
The addition of energy diffusion allows for a diffusive spread of particles with momenta both above and below the characteristic momentum. 

We consider power-law dependence for the diffusive constant, 
 \begin{eqnarray}
t_0 D_0 = D q^\alpha
 \end{eqnarray}
The following constants and variables are used to modify the diffusion term into a modified Bessel equation
\begin{eqnarray}
d &  = & - (1 + 6/\alpha) \\
y & = & \exp(\spr \alpha/\gamma) \\
z & = &  (\qpr)^{-\alpha/2 } \\
g & = & z^{(1-d)/2} h
\end{eqnarray} 
which reduces equation~(\ref{eq:g3}) into the following form:
\begin{eqnarray}
\frac{\partial h}{\partial y} - \frac{\alpha \gamma D}{4} z^{-(1+d)/2} \frac{\partial}{\partial z} \left( z^d \frac{\partial}{\partial z} \left[ z^{(1-d)/2} h \right]  \right) = 0 
\label{eq:h}
\end{eqnarray}

The solution for equation~(\ref{eq:h}) is found using a Hankel transform:
\begin{eqnarray}
H_\nu(y,k) & = &  \int_0^\infty h(y,z) J_\nu(kz) z \; dz  \\ 
h(y,z) & = & \int_0^\infty H_\nu(y,k) J_\nu(kz) k \; dk 
\end{eqnarray}
where $J_\nu$ is the standard Bessel's function of order $\nu$. With $\nu = |1-d|/2$, equation~(\ref{eq:h}) is an eigenfunction for the Bessel function, with the following Hankel transform
 \begin{eqnarray}
\frac{\partial H_\nu}{\partial y} + \frac{\alpha \gamma D k^2}{4} H_\nu  = 0
\label{eq:hankel}
\end{eqnarray}
Therefore, the transform is
\begin{eqnarray}
H_\nu(y,k) &  = &  H_\nu(y_i,k) \exp\left( -\frac{\alpha \gamma D k^2}{4}[y - y_i]  \right)  \\
& = &   \exp\left( -\frac{\alpha \gamma D k^2}{4}[y-y_i]  \right) \int_0^\infty h(y_i,z^\prime) J_\nu(kz^\prime) z^\prime \; dz^\prime 
\end{eqnarray}
where the value $y=y_i$ corresponds to the initial condition. The corresponding solution is 
\begin{eqnarray}
h(y,z) & = & \int_0^\infty k \; dk   J_\nu(kz)    \exp\left( -\frac{\alpha \gamma D k^2}{4}[y-y_i]  \right) \int_0^\infty h(y_i,z^\prime) J_\nu(kz^\prime) z^\prime \; dz^\prime  
\label{eq:hyz} 
\end{eqnarray}
The particular solution of injection at $p=p_i$ implies a delta function for $h(y_i,z) = h_i \delta(z-z_i)$. Here, $z = \sqrt{y_i}$ is the initial condition since particles are injected from $q = 1$ and $z_i = (\qpr_i)^{-\alpha/2} =  \sqrt{y_i}$. In this case, equation~(\ref{eq:hyz}) becomes 
\begin{eqnarray}
h(y,z) & = & h_i \int_0^\infty    J_\nu(kz)   J_\nu(kz_i)  \exp\left( -\frac{\alpha \gamma D k^2}{4}[y-y_i]  \right)   \;  k \; dk \\
& = &   \left(\frac{ 2  h_i}{ \alpha \gamma D}\right) \frac{1}{ (y-y_i) } \exp\left[ -  \left(\frac{ 1 }{ \alpha \gamma D}\right)  \left(\frac{ z^2 + z_i^2 }{y-y_i} \right) \right]   I_{\nu} \left[   \left(\frac{ 2  }{ \alpha \gamma D}\right)  \left(\frac{ z z_i }{y-y_i} \right) \right]
\label{eq:hyz3} 
\end{eqnarray}
Expressing the solution in terms of original variables $g(s , q)$ yields,
\begin{eqnarray}
g    & = &   \left(\frac{ 2 h_i}{ \alpha \gamma D}\right) \frac{q^{- (3 + \alpha) / 2} \exp( [s/\gamma][3 + \alpha]/2)}{ [\exp(s \alpha/\gamma) -\exp(s_i \alpha/\gamma)] } \nonumber \\
& \times & \exp\left[ -  \left(\frac{ 1 }{ \alpha \gamma D}\right)  \left(\frac{ q^{-\alpha}\exp(s\alpha/\gamma) +\exp (s_i\alpha/\gamma)}{\exp(s\alpha/\gamma) - \exp(s_i\alpha/\gamma)} \right) \right]   \nonumber \\
& \times & 
I_{\nu} \left[   \left(\frac{ 2  }{ \alpha \gamma D}\right)  \left(\frac{  q^{-\alpha/2} \exp([s+s_i]\alpha/[2\gamma]) }{\exp(s\alpha/\gamma) - \exp(s_i \alpha/\gamma)} \right) \right]   \\
 f(s,q) & = &   \left(\frac{ 2 h_i}{ \alpha \gamma D}\right) \frac{q^{- (3 + \alpha) / 2} \exp( [s/\gamma][3 + \alpha]/2)}{ [\exp(s \alpha/\gamma) -\exp(s_i \alpha/\gamma)] } \nonumber \\
 & \times & \exp\left[ -  \left(\frac{ 1 }{ \alpha \gamma D}\right)  \left(\frac{ q^{-\alpha}\exp(s\alpha/\gamma) +\exp (s_i\alpha/\gamma)}{\exp(s\alpha/\gamma) - \exp(s_i\alpha/\gamma)} \right) \right]   \nonumber \\
& \times & 
 I_{\nu} \left[   \left(\frac{ 2  }{ \alpha \gamma D}\right)  \left(\frac{  q^{-\alpha/2} \exp([s+s_i]\alpha/[2\gamma]) }{\exp(s\alpha/\gamma) - \exp(s_i \alpha/\gamma)} \right) \right]   \exp( -[s-s_i] [1+t_0/\tesc]) 
\label{eq:hyz4} 
\end{eqnarray}

By integrating over the initial time for particle injection and taking $s = 0$, we derive the steady-state distribution,
\begin{eqnarray}
f_0(q) & = & \mathrm{Lim}[s_0 \rightarrow \infty] \frac{1}{s_0} \int_{-s_0}^0 f(s = 0 , q)  \\
   & = & \mathrm{Lim}[s_0 \rightarrow \infty] \left(\frac{ 2 h_0}{ \alpha \gamma D}\right) q^{- (3 + \alpha) / 2} \int_{- s_0}^{0} ds_i \frac{1  }{ [1  -\exp(s_i \alpha/\gamma)] } \nonumber \\
 & \times & \exp\left[ -  \left(\frac{ 1 }{ \alpha \gamma D}\right)  \left(\frac{ q^{-\alpha} +\exp (s_i\alpha/\gamma)}{1 - \exp(s_i\alpha/\gamma)} \right) \right]   \nonumber \\
& \times & 
 I_{\nu} \left[   \left(\frac{ 2  }{ \alpha \gamma D}\right)  \left(\frac{  q^{-\alpha/2} \exp(s_i \alpha/[2\gamma]) }{1 - \exp(s_i \alpha/\gamma)} \right) \right]   \exp( s_i [1+t_0/\tesc]) 
\label{eq:hyz5} 
\end{eqnarray}
where $h_0 = h_i/s_0$. We treat the cases of $\alpha >0$ and $\alpha <0 $ separately. For $\alpha > 0 $, it is convenient to switch to the integration variable, 
\begin{eqnarray} 
y_i = \exp(s_i \alpha/\gamma ) 
\end{eqnarray}
in which case the steady-state solution becomes 
\begin{eqnarray}
f_0(q, \alpha > 0) 
   & = &  \left(\frac{ 2 h_0}{ \alpha \gamma D}\right) q^{- (3 + \alpha) / 2} \int_{0}^{1} \frac{dy_i}{y_i} \frac{1  }{  \alpha [1  - y_i] } \nonumber \\
 & \times & \exp\left[ -  \left(\frac{ 1 }{ \gamma D}\right)  \left(\frac{ q^{-\alpha} + y_i}{ \alpha (1 - y_i )} \right) \right]   \nonumber \\
& \times & 
 I_{\nu} \left[   \left(\frac{ 2  }{\gamma D}\right)  \left(\frac{  q^{-\alpha/2} \sqrt{y_i} }{ \alpha ( 1 - y_i ) } \right) \right]   y_i^{ [1+t_0/\tesc]  \gamma/\alpha }  
\label{eq:hyz6} 
\end{eqnarray}

The case with $\alpha < 0$ is similar to equation~(\ref{eq:hyz6}), but the limits of integration are modified,
\begin{eqnarray}
f_0(q, \alpha < 0) 
   & = &  \left(\frac{ 2 h_0}{ |\alpha| \gamma D}\right) q^{- (3 + \alpha) / 2} \int_{1}^{\infty} \frac{dy_i}{y_i} \frac{1  }{  |\alpha| [y_i - 1  ] } \nonumber \\
 & \times & \exp\left[ -  \left(\frac{ 1 }{ \gamma D}\right)  \left(\frac{ q^{-\alpha} + y_i}{ |\alpha| (y_i  - 1 )} \right) \right]   \nonumber \\
& \times & 
 I_{\nu} \left[   \left(\frac{ 2  }{\gamma D}\right)  \left(\frac{  q^{-\alpha/2} \sqrt{y_i} }{ |\alpha| ( y_i - 1 ) } \right) \right]   y_i^{ [1+t_0/\tesc]  \gamma/\alpha }  
\label{eq:hyz7} 
\end{eqnarray}
The expression is made more tractable by transforming to the integration variable $w_i = 1/y_i$, 
\begin{eqnarray}
f_0(q, \alpha < 0) 
   & = &  \left(\frac{ 2 h_0}{ |\alpha| \gamma D}\right) q^{- (3 + \alpha) / 2} \int_{0}^{1} dw_i \frac{1 }{  |\alpha| [1- w_i ] } \nonumber \\
 & \times & \exp\left[ -  \left(\frac{ 1 }{ \gamma D}\right)  \left(\frac{ w_i q^{-\alpha} + 1}{ |\alpha| (1  - w_i )} \right) \right]   \nonumber \\
& \times & 
 I_{\nu} \left[   \left(\frac{ 2  }{\gamma D}\right)  \left(\frac{  q^{-\alpha/2} \sqrt{w_i} }{ |\alpha| ( 1 - w_i ) } \right) \right]   w_i^{ [1+t_0/\tesc]  \gamma/|\alpha | }  
\label{eq:hyz8} 
\end{eqnarray}
We take, for simplicity and clarity, the escape time inversely proportional to energy such that 
\begin{eqnarray}
\frac{t_0}{\tau_\mathrm{esc}} = \sqrt{\frac{E}{E_\mathrm{esc}}}. 
\end{eqnarray} 
These equations can now be used to compare to results of  STAT and EPREM. Our analytic solutions enable us to validate our understanding of the acceleration processes seen in particle acceleration at plasma flows through separatrix layers. 

\section{Escape from the separatrix layer}
\label{sec:escape}

The paper shows the effects of particle acceleration from EPREM simulations. Particles are accelerated as the result of a  second-order acceleration process. However, the acceleration region is limited in size by the extent of separatrix layers associated with strong gradients in the magnetic field. As a result, the acceleration process has similarities with an important property of diffusive shock acceleration. The spatial localization of these processes leads to an important consideration concerning particle escape. Specifically, there is a high-energy threshold above which particles cannot be contained near the site of acceleration. Above the escape energy threshold, particles efficiently leave the acceleration site, which results in a roll-over in the accelerated energy spectrum.  

Results of \cite{Schwadron:2015SEP} reveal that escape from the acceleration region near shocks and compressions must be considered when treating the behavior of distribution functions at relatively high energy. Particle escape from an accelerator occurs in a number of different ways. In \cite{Schwadron:2015SEP}, plasma flows along a shock surface sweep magnetic field lines out of the acceleration region at a steady rate. This form of escape is particularly important at shocks and compressions where the magnetic field is almost perpendicular to the plasma velocity gradient, which acts as the source of particle acceleration. 

\textcolor{black}{In separatrix layers in the low corona on open magnetic field lines, plasma flows along the magnetic field. Streamlines remain fixed and tied to any QSL structures they pass through.  The magnetic field remains tied to the acceleration region over long periods, and particles escape by leaving the accelaration site diffusively.  The escape process occurs because the scattering mean free path is relatively small immediately outside the separatrix layer (see panel 4 in Figures 5 and 6). However, the scattering mean free path grows  progressively beyond $\sim$ 4 R$_\odot$ where energetic particles are able to stream outward  with a scattering mean free path $> 0.09$ R$_\odot$. Because the energetic particle fluxes are larger closer to the Sun, and the scattering mean free path increases with distance beyond 4  R$_\odot$, there is a strong gradient in the fluxes, and an outward streaming that reduces the probability of return fluxes. The top panel  in Figures 5 and 6 show the strong outward reduction in energetic particle fluxes, which indicates the escape process at work.  }

 \textcolor{black}{ The EPREM model captures the escape process  since changes in the scattering mean free path are treated  explicitly. The analytical model however assumes a fixed mean free path and escape is treated as a loss process with an escape time $\tau_\mathrm{esc}$.   }

 Scattering mean free paths are in the corona than at 1 au. For example, \cite{Droge:2000} find typical scattering mean free paths in the MeV/nuc range in the range of 0.02 to 0.5 au observed near 1 au.  We have extrapolated the radial dependence of the scattering mean free path (MFP) by scaling the field inversely to the power 3/4, $\lambda_\parallel \propto B^{-3/4}$. With a field falling as $B\propto 1/r^2$, and a 0.1 au MFP at 1 au, we find a MFP of $9\times 10^{-5}$ au (or 0.02 $R_{\odot}$) at 2 $R_{\odot}$ outside the separatrix layer. 

The scaling for the MFP is also in line with recent work from the IS$\odot$IS instruments \cite[]{McComas:2016} on Parker Solar Probe \cite[]{Fox:2016}. 
\cite{Giacalone:2023} studied a shock observed  for 0.05 to 2 MeV protons when the spacecraft was near 0.35 au. Particles far upstream from the shock showed scattering mean free paths in the range 0.02 to 0.1 au (4 - 21 $R_{\odot}$), which is compatible with typical ranges observed at 1 au. It is notable that near the shock, increased levels of turbulence reduce the scattering mean free path by a factor of $\sim 1/10$, which greatly increases the rate of acceleration near the shock. As done previously for the low corona, we scale a 0.2 au MFP at 1 au to PSP's position 0.35 au (75 $R_{\odot}$), and find a MFP of $0.04$ au, which is roughly the average MFP found by  \cite{Giacalone:2023} near PSP far upstream from the shock. 

In the analytic model, we estimate the escape time $\tau_\mathrm{esc} = L^2/\kappa_\parallel$. With a constant  mean free path,  
\begin{eqnarray}
\tau_\mathrm{esc} & = &  t_0 \sqrt{\frac{E_\mathrm{esc}}{E}},
\end{eqnarray} 
where 
\begin{eqnarray}
E_\mathrm{esc} & = &  (3 L/\lambda_\parallel)^2 E_\mathrm{sw},
\end{eqnarray}
$t_0 = L/u_{su}$ is the transit time through the separatrix layer based on the upstream plasma speed $u_{su}$, and $L$ is the spatial scale over which diffusion proceeds within the  separatrix layer along the mean field line.
The quantity $E_\mathrm{sw} = m u_{su}^2/2$ is the characteristic solar wind energy for a species of mass $m$. With $\lambda_\parallel = 0.02$ $R_{\odot}$, $L=0.5$ $R_{\odot}$ ,  and $u = 70$ km/s near the separatrix layer, we find $E_\mathrm{esc} \approx 0.2$ MeV/nuc. 

The escape process is particularly important for thin separatrix layers. It is interesting to note that QSLs and pseudo-streamers often show thicker separatrix layers that exceed 1.2 $R_{\odot}$. It is precisely in these circumstances that the escape energy increases (with roughly the square of the characteristic separatrix thickness). This, in addition to increasing the magnetic pumping effect \citep[e.g.,][]{fisk:2010} increases both the fluxes and particle energies from QSLs. 

\section{Waiting Time Distributions and Stochastic Processes}
\label{sec:waiting}
The example of waiting time distributions provides perhaps a simplified way to understand the connection between acceleration at QSLs and stochastic processes. Suppose there is a series of events separated by waiting time $t$. These events may be of any sort provided that they are randomly distributed. For example, previous treatments of waiting time distributions have been applied to solar flares \cite[]{Wheatland:2000, Dragulescu:2000, Yakovenko:2009}. A given inverse exponential distribution of waiting times has a mean event rate $\beta$. The probability of an event with a given waiting time is therefore,
\begin{eqnarray}
P_\tau(t) = \beta \exp( - \beta t). 
\label{eq:P1}
\end{eqnarray}
We can take the first moment of this probability distribution to determine the average waiting time, 
\begin{eqnarray}
\tau = \int_0^\infty t P(t)\,dt  = 1/ \beta. 
\label{eq:P2}
\end{eqnarray}
The average waiting time is simply the inverse of the mean event rate.  This formulation is adequate provided that there is only one process proceeding with a specific mean event rate. 

\cite{Wheatland:2000} developed the more general case when there is an ensemble of processes each with a distinct occurrence rate. The processes are treated as states, and have random or Poisson behavior if the states maximize the entropy associated with these states. Applying this entropy constraint to ensure the states are randomly distributed results in a waiting time distribution that conforms to a kappa distribution.

These same considerations applied to waiting time distributions also apply to the distribution functions of energetic particles \cite[]{Schwadron:2010h}. The fact that the waiting time distribution for a given process depends on the inverse of an exponential in time aligns precisely with behavior of a given acceleration process, which is exemplified in Appendix \S A.2 (see equation~(\ref{eq:exptime2})).  

The waiting time has another connection to stochastic processes when considering the spatial distribution of particles from accelerated sources. Consider a characteristic distance $L_s$ from a source defined such that $L_s = v /\beta$. With this definition, the waiting time distribution is expressed
\begin{eqnarray}
P_\tau(t) = (v/L_s ) \exp( - v t / L_s ). 
\label{eq:P3}
\end{eqnarray}
The average waiting time is now 
$\tau   = L_s/ v $ and represents the characteristic propagation time from the source. Observing particles with small waiting times $t < L_s/v$ is more likely because  particles build-up near the source and are separated over distance $\ell_s < L_s$. In contrast, observing particles with large waiting times $t > L_s/v$ is less likely since they have separations $\ell_s > L_s$ larger than the distance to the source. 

We arrive at a similar form for the waiting time distribution by accounting for particle acceleration.  In the presence of acceleration, particle speeds increase with time.  Shorter distances between particles indicates short waiting times, a location nearer the source, and smaller amounts of acceleration.  Particles propagating further from the source are capable of experiencing larger amounts of acceleration.  

The accelerators could be the presence of small-scale fluctuations of the magnetic field magnitude and associated transit-time damping.   For example, particles can  cross field lines and gain energy from the convective electric field, $\mathbf{u} \times \mathbf{B}$. In this example, the acceleration is applied over the waiting time, and the speed increases as $v = c_b +  a t$, where $a$ is the average acceleration over time $t$, and $c_b$ represents the characteristic speed for the bulk of the distribution. Larger waiting times are associated with a minority of particles capable of accessing larger amounts of acceleration. The waiting time distribution is  expressed as
\begin{eqnarray}
P_\tau(v) = \beta \exp( - \beta [v-c_b]/a ). 
\label{eq:P4}
\end{eqnarray}
The break in the distribution occurs at a characteristic speed  $v = a /\beta$. The speed distribution, $P_u(v)$, follows, 
  \begin{eqnarray}
P_u(v) & \propto &   (1/[dv/dt]) P_\tau(v) \nonumber \\
   & = &    C_n (\beta /a)  \exp( - \beta [v-c_b]/a ) \nonumber \\
   & = &   (\beta/a) \exp( - \beta v/a )
\label{eq:P5}
\end{eqnarray}
Here, the  normalization constant $C_n$ is determined by the definition of a probability distribution, $\int_0^\infty dv P_u = 1$, while the mean speed of the distribution is given by $\langle v \rangle \equiv u = a /\beta$.

 \cite{Schwadron:2010h} considered the waiting time distribution as a random or Poisson process. 
The almost equal probability
of waiting times below the break at $t < 1/\beta$ qualifies the Poisson process as random. In an
acceleration process, it also takes some time to achieve a given
particle speed $v$. The acceleration time behaves effectively as the waiting time. 

Each process is associated with a  probability distribution and a mean speed $u$ tied to the process rate. An array of these processes is considered as an array of states in a system with a given mean inverse speed $\zeta_0$. Since the acceleration processes are random, the distribution of states is subject to the maximization of the Boltzmann entropy, given the constraint of specific average inverse speed  \cite[]{Schwadron:2010h}. Therefore, the state distribution is given by, 
\begin{eqnarray}
F(\zeta ) = \exp[-\zeta  / \zeta_0 ] / \zeta_0 .
\end{eqnarray}
\cite{Schwadron:2010h} demonstrated that the superposition of these states results in a kappa-distribution. The superposed probability distribution is 
\begin{eqnarray}
P_s(v) = 2 \zeta_0 /( v\zeta_0 +1 )^3
\end{eqnarray}
and the associated particle distribution is 
\begin{eqnarray}
f(v > c_s) & = & n P_s(v)/(4 \pi v^2) \nonumber \\
 & = & \left( \frac{n}{2\pi v^2}  \right) \frac{ \zeta_0}{\left( v \zeta_0 + 1\right)^3} 
\end{eqnarray}
where $n$ is the density of the distribution, and the distribution function is explicitly considered for particle speeds greater than the core or thermal speed of the distribution, $v > c_s$.  

\end{document}